\documentclass[11pt]{article}
\usepackage{algorithmicx,  algorithm, algpseudocode,amsmath, amsthm, amssymb, authblk, bm,caption, datatool, enumerate, float, glossaries, graphicx, graphics, hyperref, index,  listings, multirow, nomencl, natbib,ltablex, longtable,  rotating, scalerel,  setspace, subfig,tabularx, titlesec,url, verbatim, 
}
\usepackage[table,xcdraw]{xcolor}
\usepackage[a4paper,inner=2cm,outer=2cm,top=2.5cm,bottom=2.5cm,pdftex]{geometry}

\usepackage[english]{babel}
\hypersetup{
    colorlinks,
    linkcolor={blue!80!black},
    citecolor={blue!80!black},
    urlcolor={blue!80!black}
}
\graphicspath{ {Figures_paper/} }
\algnewcommand{\LineComment}[1]{\State \(\triangleright\) #1}

\title{Multiple Imputation Approaches for Epoch-level Accelerometer data in Trials }
\author{Mia S. Tackney  \textsuperscript{$\dagger$} \thanks{Address correspondence to Mia S. Tackney, MRC Biostatistics Unit, University of Cambridge, East Forvie Site, Robinson Way, Cambridge CB2 0SR, UK; E-mail: Mia.Tackney@mrc-bsu.cam.ac.uk} , 
 Elizabeth Williamson\thanks{Department of Medical Statistics, London School of Hygiene and Tropical Medicine, UK},
 Derek G. Cook\thanks{Population Health Research Institute, St George's, University of London, UK} ,
 Elizabeth Limb \textsuperscript{$\ddagger$},
 Tess Harris \textsuperscript{$\ddagger$} ,
 James Carpenter \textsuperscript{$\dagger$} \thanks{MRC Clinical Trials Unit at University College London, UK}}

\begin{document}
\maketitle

\begin{abstract}
Clinical trials that investigate interventions on physical activity often use accelerometers to measure step count at a very granular level, often in 5-second epochs. Participants typically wear the accelerometer for a week-long period at baseline, and for one or more week-long follow-up periods after the intervention. The data is usually aggregated to provide daily or weekly step counts for the primary analysis. Missing data are common as participants may not wear the device as per protocol. Approaches to handling missing data in the literature have largely defined missingness on the day level using a threshold on daily wear time, which leads to loss of information on the time of day when data are missing. We propose an approach to identifying and classifying missingness at the finer epoch-level, and then present two approaches to handling missingness. Firstly, we present a parametric approach which takes into account the number of missing epochs per day. Secondly, we describe a non-parametric approach to Multiple Imputation (MI) where missing periods during the day are replaced by donor data from the same person where possible, or data from a different person who is matched on demographic and physical activity-related variables. Our simulation studies comparing these approaches in a number of settings show that the non-parametric approach leads to estimates of the effect of treatment that are least biased while maintaining small standard errors. We illustrate the application of these different MI strategies to the analysis of the 2017 PACE-UP Trial. The proposed framework of classifying missingness and applying MI at the  epoch-level is likely to be applicable to a number of different outcomes and data from other wearable devices. 
\end{abstract}


\section{Introduction} 
Wearable devices are increasingly becoming popular tools to measure health outcomes in clinical trials. In trials that investigate interventions aimed to increase physical activity, accelerometers have been used in a number of studies to evaluate impact on participants' step count \citep{Harris2015, Harris2017, Harris2018, Ismail2019}. Accelerometers measure acceleration in three dimensions in very fine intervals of time, typically in 5-second intervals or epochs, and offer a more objective measure of physical activity with reduced participant burden compared to self-report approaches. Outputs of interest from accelerometers include vector magnitude (VM), which summarizes the accelerations in three dimensions, step count and time spent in different physical activity intensities (e.g. sedentary, light, moderate-to vigorous physical activity) \citep{LeegerAschmann2019}. Missing data can occur in a number of ways in this setting; for example, there may be device failure due to the battery running out or water damage, or participants may remove or forget to wear the accelerometer for periods of time during the day. Analyses should account for the missingness in a suitable way in order for results to be unbiased and to reflect the uncertainty appropriately. Multiple Imputation (MI) is a flexible and powerful approach to handling missing data, and has previously been applied to the accelerometer setting where outcomes are aggregated at the day level \citep{Tackney2021}.

\begin{figure}[H]
\centering
\includegraphics[scale=0.44]{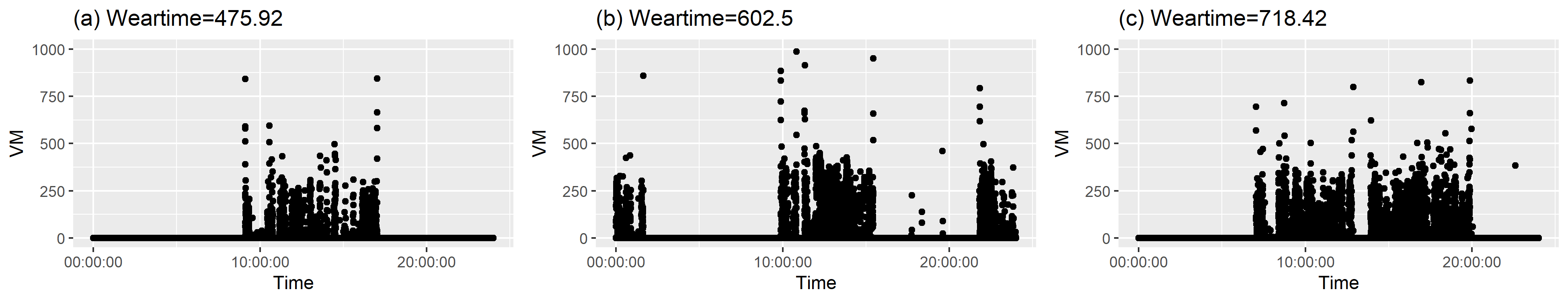}
\caption{Vector Magnitude (VM) is plotted against time for data from three days from three different individuals from the PACE-UP trial. }
\label{examples}
\end{figure}

Approaches which apply MI to day-level step counts require missingness to also be determined at the day-level. A popular approach in the literature is to define a day as missing if a participant wore the device for less than 540 minutes in a day \citep{Harris2015, Harris2017, Harris2018, Ismail2019}. Other common choices of threshold include 360 minutes of wear time \citep{DeCraemer2016} and 600 minutes of wear time \citep{Goode2015, Cameron2017}. Defining missingness at the aggregate day-level has some drawbacks. Thus participants may provide valuable data on so-called ``missing days" (e.g. days with less than 540 minutes of wear time) which would then become discarded; for example, Figure \ref{examples} plot (a) displays VM from a day where the device was worn for 475.92 minutes, which is slightly short of the required weartime. Equally, participants who do provide at least 540 minutes of weartime could potentially still have missing parts of days; for example, Figure \ref{examples} plots (b) and (c) are examples of days where weartime is adequate, but there are periods during the day where no data is recorded and could potentially be missing. \cite{Tackney2021} proposed an alternative approach where days are classified as  missing, partially observed or observed, and partially observed days are treated as right-censored data, which retains the information from days where participants provide some, but insufficient, data. However, even with this approach, information on times of day where missingness takes place is discarded. Examining the time of day when participants have missing data could be valuable in trying to restore information and can lead to greater clarity. Applying MI at the epoch-level have had limited attention in the literature so far.

 Multiple Imputation (MI) is a flexible and powerful approach to handling missing data. MI accounts for uncertainty in the missing values, and allows for sensitivity analysis to explore departures from the Missing at Random (MAR) assumption. Further, MI can allow for outcomes to be on an aggregate level (day- or week-level), while missingness is handled on the finer epoch-level.  To date, there has been limited exploration of missingness on the epoch-level data in the literature. \cite{Lee2018} propose a zero-inflated Poisson and log-normal mixture distribution which allows for imputations at the epoch-level.  \cite{Butera2019} proposed a non-parametric (hot deck) approach to MI at the epoch-level. When data are MAR, their simulation studies showed that the non-parametric approach produced less bias and improved coverage compared to available case and complete case approaches. However, neither of these epoch-level approaches to imputation fully reflect the complexity of defining missingness at the epoch-level. \cite{Lee2018} assume a window of between 9am and 9pm in which participants are awake, and define missingness as intervals of at least 20 minutes of no recorded acceleration. Limiting the window of data considered between 9am and 9pm ignores the variation within and between people's waking and sleeping times, and does not acknowledge the possibility that some of these periods may actually be due to the participant removing their device during sleep, which is per-protocol and should not be imputed. In their simulation studies, \cite{Butera2019} induce missingness in two-hour blocks of time, which simplifies the complexity of missingness in genuine datasets. 

This study aims to characterize the common patterns of missing accelerometer data at the epoch-level, and handle epoch-level missingness utilising MI using both parametric and non-parametric approaches, illustrated by a specific trial example. We first describe the \nameref{PACE-UP}, and then describe \nameref{MI} and the challenges in its application in the accelerometer context. We then describe the non-parametric and parametric approaches in \nameref{Proposal}. These approaches are validated through \nameref{Sim}, and their performances are compared in an  \nameref{App}.

\section{PACE-UP trial}
\label{PACE-UP}
We illustrate the missing data issues in the accelerometer context using the 2017 PACE-UP trial as a motivating example. The PACE-UP trial investigated postal and nurse-supported interventions for increasing physical activity in patients aged 45-75 years from seven primary care practices in London. Randomization was by household (to avoid couple contamination) and block randomization was used within seven primary care practices. Of the 1023 patients in the trial, 338 were randomized to usual care, 339 to postal pedometer intervention and 346 to nurse-supported intervention. The participants were provided with an ActiGraph GT3X accelerometer (ActiGraph, FL, USA) for a period of seven
consecutive days on four separate occasions, which we refer to as \textit{time points}: baseline, 3 months, 12 months,
and 3 years. They were instructed to wear the accelerometer on the hip using a belt during waking hours, except when swimming or showering. The protocol and results of the trial have been reported previously \citep{Harris2017, Harris2018}, and the trial showed that physical activity increased in both intervention groups compared to usual care.

In the reported trial results, days are defined as missing if wear time is less than 540 minutes \citep{Harris2017}. Further, at least five non-missing days at baseline, and at least one non-missing day at 12 Months are required to be included in the primary analysis. Of the 1023 patients who were randomized, $93\%$ of participants were included in the 12-month primary analysis. The average of the non-missing days were computed at baseline and 12 months to assess change in step count, adjusting for day of week, and day-order-of-wear. The primary analysis assumes that the data are missing at random (MAR), and sensitivity analyses were conducted to assess the impact of using different thresholds on weartime for defining missingness, and to assess the impact of data being missing not at random (MNAR) \citep{Harris2017}.

\section{Multiple Imputation}
\label{MI}

Multiple imputation (MI) is a flexible and practical approach to the analysis of datasets with missing values. An imputation model is specified, which is a model for the posterior predictive distribution of the missing outcomes given the observed data \citep{Harel2007}. This model is used to impute missing data with $M$ plausible values, resulting in a total of $M$ sets of complete data. The analysis model is fitted to each of the $M$ datasets, and the results for inference are combined using Rubin's rules \citep{Rubin1976} to take full account of the uncertainty due to the missing values. If the imputation model is specified appropriately, MI provides valid and efficient inference under the assumption that the data are MAR given the observed data in the imputation model \citep[Ch.2]{Carpenter2012}. Sensitivity analysis to assess the robustness of the results to missing data assumptions is recommended \citep{Cro2020}. An attractive feature of MI in the accelerometer setting is that the imputation model and the analysis model are separate, which allows missingness to be defined on a different level than the level specified in the analysis model. For example, the analysis model may have as the outcome the step counts averaged across the week and the imputation model may handle missingness at the finer day- or epoch-level to achieve more precise imputations.

The imputation model is typically specified as an explicit parametric model for the predictive distribution of the missing variables given the observed data. For example, a multivariate normal model can be specified, or Tobit regression may be used in the accelerometer setting to incorporate step counts as right-censored observations in the imputation model if participants took an insufficient number of steps. A parametric imputation model may include additional auxiliary variables which are not in the analysis model, but are predictive of missingness or step count. The inclusion of variables such as daily weather variables in the accelerometer setting can help to make the MAR assumption more plausible \citep{Tackney2021}. An alternative approach is to use non-parametric or hot-deck imputation, which replaces missing values with donor data, which are observed data that have been identified to be of similar characteristics to the missing data \citep{Andridge2010}. An important advantage of this approach is that it is compatible with complicated relationships in the dataset which do not have to be specified via a statistical model \cite[p. 181]{Carpenter2012}. 

Accelerometer data presents two main challenges which need to be addressed before MI can be directly applied; firstly, the challenge of defining missingness at the epoch-level, and secondly, the challenge of handling difficult distributions.

\subsection{Complications for MI in the accelerometer context}

\subsubsection*{Challenge 1: Defining missingness at the epoch-level\\} 

The first challenge in defining missing data in the accelerometer context is identifying when participants have removed the device. Unlike more sophisticated wrist-worn devices, the GT3X+ accelerometers did not measure pulse or heart rate. It is difficult to decide whether a period with no movement recorded is: (A) a period where the participant is wearing the device but staying still, (B) a period where the participant has removed the device per protocol, such as during sleep, or (C) whether the participant has removed the device during the day and is an instance of missing data. Participants' movement is often quantified by Vector Magnitude (VM), which is the square root of the sum of the acceleration in each component squared. Figure \ref{examples} displays plots of VM against time for three days from different participants in the PACE-UP trial. In plot (c), there is a short period of no activity in the middle of the day, which could be missing data, or perhaps could be the participant lying still. In contrast, the longer periods of no movement in the morning and in the evening are very likely to be the participant removing the device for sleep, as per protocol. There is a need to classify these types of activities at the epoch-level, in order to identify the missing intervals that need to be imputed through MI.

\subsubsection*{Challenge 2: Difficult distributions\\} 

Accelerometer data at the epoch-level are characterized by a large proportion of zeros, a heavy positive skew, and high autocorrelation \citep{Lee2018}. The complexity in the distribution of epoch-level data means that parametric approaches to imputation that assume a standard distribution, such as the normal distribution, are likely to be inappropriate. 

Further, epoch-level accelerometer data collected over the course of a week is characterized by complex within-person patterns. There are patterns of activity that are dependent on time of day, and these patterns are generally different on weekends compared to weekdays. Allowing for these patterns on the epoch-level is a statistical and computational challenge.

\section{Proposed Approaches}
\label{Proposal}

\subsection*{Classifying Missingness at the Epoch Level}

\label{Classificiation}
Resolving the first challenge of identifying missing periods begins by identifying \textit{zero-count} periods. Zero-count periods are intervals of time where VM is continuously zero over a specified threshold, usually set at 20, 60 or 90 minutes, where it can be assumed that the device is removed \citep{Evenson2009}. Some authors recommend allowing for a \textit{spike tolerance}, that is, allowing for an interval of up to 2 minutes of non-zero VM to account for inadvertent movements of the device, such as the device being moved across the table \citep{Choi2011}. We adopt the definition used in the PACE-UP trial, where a 60 minute threshold was used, allowing for a 2-minute spike tolerance. We note that zero-count periods are sometimes referred to as \textit{non-wear} in the literature, but we reserve \textit{non-wear} to refer more specifically to periods where people are likely to have removed the device. 

\begin{figure}
\center
\includegraphics[scale=0.44]{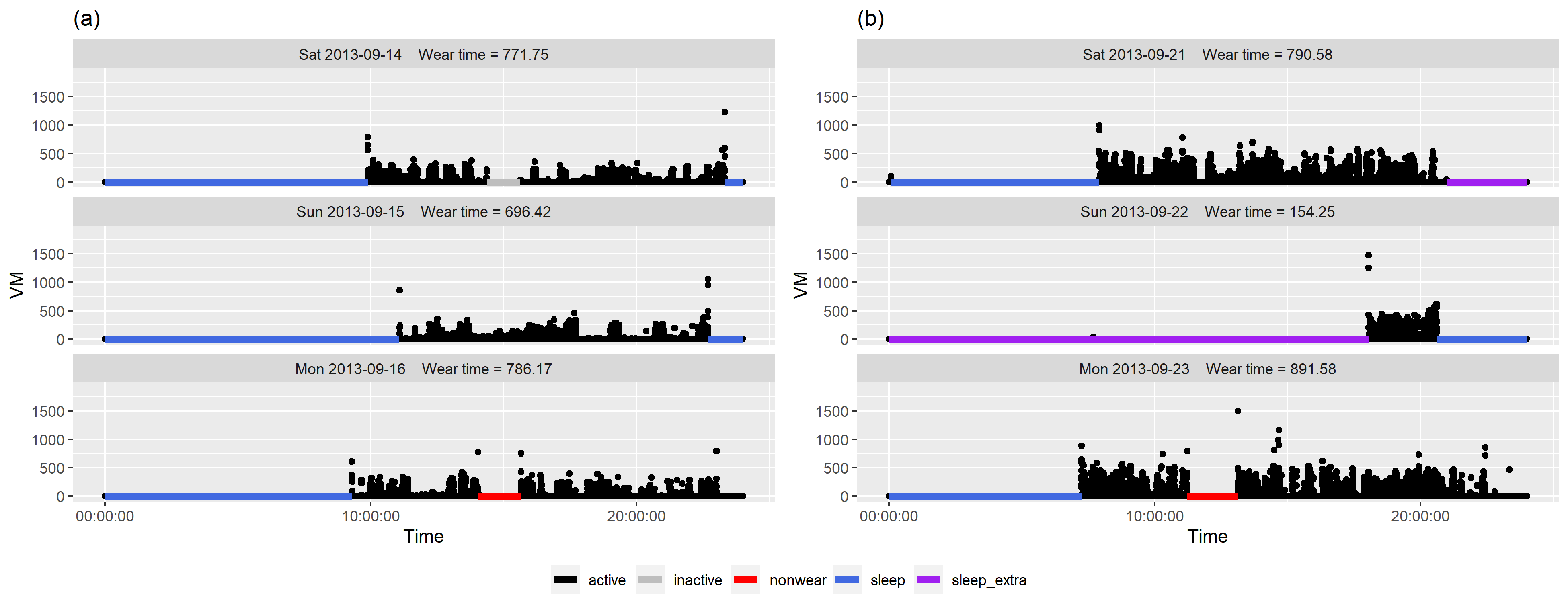}
\caption{Vector Magnitude is plotted against time for two individuals. In (a), we observe an \textit{inactive} period on Saturday and a \textit{nonwear} period on Monday. In (b), we observe three days, where the first and third days display sufficient wear-time, but on the second day, the accelerometer was not worn for the most part of the day, resulting in a \textit{sleep-extra} period. }
\label{missingness_types}
\end{figure}

Zero-count periods include periods where participants are wearing the device, but are staying still. In order to distinguish these periods, we note that putting on and removing the device requires a sharp movement, which is detectable as a spike in VM. Empirically, we explored the data and confirmed that this is the case; VM greater than 600 is typically incurred when the accelerometer is put on or removed. We classify zero-count periods lasting between one and five hours with VM of at least 600 in the 2-minute interval before or after the period, as \textit{non-wear} periods. For example, in Figure \ref{missingness_types} panel (a), we observe on Monday a zero-count period indicated in red where the high VM points immediately before and after indicate that the device was removed, so we classify this as non-wear. In contrast, on Saturday, we observe a zero-count period, indicated in grey, where no high VM is detected before or after. In this case, it is possible that the participant is still wearing the device, but staying very still. We classify zero-count periods of up to 3 hours, where no high VM are detected, as \textit{inactive} periods, which are not treated as missing. 

Missingness also occurs when people put on the device later in the day than when they are expected to wake up, or remove the device earlier than when they are expected to go to bed. This is visible by a very long period zero-count period which would include the time when the participant is expected to be asleep. For example, in Figure \ref{missingness_types} panel (b), the purple area illustrates a case where the participant has not worn the device until late in the evening. We refer to these extended zero-count periods, lasting longer than 15 hours, as \textit{sleep-extra} periods. Based on these observations, we classify zero-count periods into: 
\begin{itemize}
\item \textit{Inactive}: a shorter continuous zero-count period, lasting between 1 and 3 hours, where no high VM is detected (VM does not exceed 600 in the 2 minutes just before/after the zero-count period). This suggests that accelerometer is still worn by the individual, but they are staying very still. During a period of 1 to 3 hours, it is plausible that a person is staying still. Inactive periods are not missing periods. 
\item \textit{Non-wear}:  a continuous zero-count period (lasting between 1 and 5 hours) where VM exceeds $600$ in the 2 minutes just before or after the period. This suggests the accelerometer has been taken on/off.  Further, any zero-count period between 3 and 5 hours is classified as non-wear. It is less plausible that a person could stay still for an extended period of time; experience with using the accelerometer suggests that it is very unlikely that it will register no movement for over 3 hours if the device is being worn. Non-wear periods are missing periods. 
\item \textit{Sleep}: A zero-count period lasting between 5 and 15 hours. Sleep periods are not missing periods. 
\item \textit{Sleep-extra}: A zero-count period lasting longer than 15 hours. Such an extended period of sleep suggests that a person delayed putting on the device in the morning, and/or took it off too early in the evening. Sleep-extra periods contain missing periods. 
\end{itemize}

Examples of plots displaying  Vector Magnitude (VM) against Time for the 7-day period at baseline and 12 months for specific patients are shown in the Appendix \nameref{Epoch-plots}, with zero-count periods classified. 

Using this classification, days/participants who do not have non-wear or sleep-extra periods are considered fully observed. Non-wear and sleep-extra periods lead to missing periods that need to be accounted for in the analysis. For non-wear periods, the start and end times of the missing period are equal to the start and end times of the non-wear period. However, for sleep-extra periods, which include pro-protocol sleep periods which are not considered missing, the start and/or end times need to be estimated. If the sleep-extra period for a participant falls on a weekday, the \textit{average sleep window} for weekdays is computed by taking weekdays from this participant with completely observed data, and finding the interval between the average time at which the participant goes to sleep, and the average time at which the participant wakes up.  The missing intervals consist of sleep-extra period, minus period which lies in the average sleep window. If the missing interval falls on a weekend, then, if the other weekend day is fully observed, the average sleep window of that weekend day is used to compute the missing intervals. If both weekend days have missingness, the average sleep window is obtained from the weekdays, adding an empirically-based estimate of the average shift in waking times at the weekend compared to weekday. Since this was approximately an hour in these data, we rounded it to exactly one hour for convenience.

Having defined missing periods at the epoch-level, we wish to handle missingness with Multiple Imputation. We describe two approaches to overcoming the second challenge of epoch-level data having complex distributions: a parametric approach, described in  \nameref{Para approach} and a non-parametric approach, described in  \nameref{Non-para approach}. We introduce some notation to describe the approaches. We denote by $y_{i, j, k, l}$ the step count for patient $i$, at timepoint $j$, on day $k$ and epoch $l$, and we denote by $y_{i, j, k, l_p:l_q}$ the step counts over an interval between epoch $l_p$ and epoch $l_q$, where $l_p < l_q$.  We assume, without loss of generality, that data are recorded in 5-second epochs. We denote by ${y}_{i, j, k, l_p:l_q}^{obs}$ an interval that is observed, and ${y}_{i, j, k, l_p:l_q}^{mis}$ an interval that is missing. We denote by ${y}_{i, j, k,.}$ the day-level step counts for day $k$, and $\bar{y}_{i, j,.,.}$ the mean of the daily-level step counts for timepoint $j$.

\subsection{Parametric approach}
\label{Para approach}

In the parametric approach to MI, in order to overcome difficulty of epoch-level step count data having a high proportion of zeros and extreme positive skew, we aggregate the data to the day level. Day-level step counts still have a positive skew, but this can be handled with a log-transformation to help make the normality assumption plausible. A parametric approach to MI at the day-level was proposed in \cite{Tackney2021}; we make a crucial adaptation to this approach to incorporate information about missingness at the epoch-level.

The day-level approach in \cite{Tackney2021} classified step counts as completely observed if weartime $\geq 540$, partially observed if $0 < \mbox{weartime} < 540$, and missing if weartime $=0$. In our adapted approach, we move away from using a threshold based on weartime and take into account the missing periods detected at the epoch-level. We consider a daily step count as completely observed if there are no missing periods. A daily step count is partially observed if there are non-wear or sleep-extra periods, and completely missing if no data was recorded. 

We consider the daily step counts as right-censored data if they are partially observed or missing, and use Tobit regression to conduct the imputation. Tobit regression requires specification of lower and/or upper bounds for each observation. For days that are completely observed, the lower and upper bound are the recorded logged step counts. For days that are completely missing, the lower bound is zero, and the upper bound is a value higher than the highest observed logged daily step count in the data (e.g., 10.5 on the log scale). For days where some activity is observed, but which have missing periods, the lower bound is the recorded step count, and we propose a \textit{Person-specific upper bound}, calculated as $\log( {y}_{i, j, k,.} + 5 \lambda_{i,j,k})$, where $\lambda_{i,j,k}$ is the number of missing epochs for participant $i$ at timepoint $j$ for day $k$. This assumes that the upper bound of total step count would allow up to 1 step each second (5 steps per epoch) in the missing period. This approach adjusts the upper bound according to the quantity of missingness detected at the epoch-level. We compare using the person-specific upper bound to using a \textit{Generic upper bound}, which sets the upper bound as 10.5 on the log scale. This generic upper bound was used in \citep{Tackney2021} and serves as a comparison with previously suggested approaches.

We assume that the logged daily step counts are jointly normally distributed, possibly dependent on baseline characteristics such as sex, age and BMI, and further, we assume that the data are MAR. Activity patterns across days are accounted for through adopting a joint model for the logged daily step counts, and through the addition of covariates. We impute separately within each arm. After imputation, the log of the daily step counts are exponentiated. The $M$ complete datasets on the step count scale can then be analysed separately, and combined using Rubin's rules.

\subsection{Non-parametric approach}
\label{Non-para approach}

Secondly, we consider a non-parametric approach to using MI at the epoch-level. Instead of specifying a parametric statistical model for the distribution of the missing data given the observed data, a non-parametric approach proceeds by imputing missing periods with observed periods from the same time of day, from the same participant, but from a different day of the week, where possible. If this is not possible, the interval is imputed from a different participant who is as similar as possible according to demographic metrics.

We assume that imputation is within a treatment arm, and within a specific time interval $j$. For each participant $i$, missing periods are identified and classified. If any missing period is spread between two days, for example between epoch $l_p$ on day $k$ and epoch $l_q$ on day $k+1$, this is split into two missing periods, $\bm{y}_{i, j, k, l_p: 17280}^{mis}$ and $\bm{y}_{i, j, k+1, 1:l_q}^{mis}$. We obtain the set of missing intervals $\mathcal{I}_i$ for participant $i$. We denote by $\mid \mathcal{I}_i \mid$ the size of the set $\mathcal{I}_i$. 

Suppose $\bm{y}_{i, j, k, l_p:l_q}^{mis}$ is the $g$th  missing period in a non-empty set of missing periods for participant $i$, $\mathcal{I}_i$, where $g \in \left\{1, ..., \mid \mathcal{I}_i \mid \right\}$. Imputation proceeds as follows: 

\begin{itemize}
\item obtain the self-donor pool $\mathcal{SD}_{i, g}$, which consist of observed intervals $\bm{y}_{i, j, k', l_p:l_q}^{obs}$, where $k \neq k'$. 
\item If $\mid \mathcal{SD}_{i, g} \mid > 4$, sample $M$ times with replacement from $\mathcal{SD}_{i, g}$ with equal sampling probabilities, to obtain $M$ imputed intervals. 
\item If $\mid \mathcal{SD}_{i, g} \mid \leq 4$, we obtain a non-self donor from the pool of participants who do not have missing periods that overlap with the $g$th missing period of patient $i$. From this pool of participants, we match perfectly on sex, and compute the Mahalanobis distance between patient $i$ and and the participants in the non-self donor pool for BMI and age. If imputing at 12 Months, the average step count at baseline and average weartime at baseline can additionally be included in the Mahalanobis distance. We note that imputation is conducted within a specific time point, so even if baseline characteristics are used to compute a Mahalanobis distance, data used in the imputation is from 12 Months. We then compute sampling weights for each participant in the donor pool by taking the inverse of the Mahalanobis distance for that participant as a fraction of the sum of inverses of Mahalanobis distances from all participants in the donor pool. One donor is selected using these sampling weights. From this selected donor $i'$, seven observed intervals $\bm{y}_{i', j, k, l_p:l_q}^{mis}$ for $k \in \left\{1, ..., 7 \right\}$ are identified. By sampling from these seven intervals with replacement, $M$ imputed intervals are obtained. 
\end{itemize}

At 12 Months, there may be a small number of participants with five or more days with wear time $< 300$ minutes. In this case, there is insufficient data to compute the participant's sleep-window and to identify the intervals that need to be imputed. While this does not occur at baseline (for example, in the PACE-UP trial, participants needed to provide at least five days with wear time of at least 540 minutes at baseline to be included in the study), this can occur at 12 Months. Here, we discard data from this participant and impute the entire week with the pool of donors who provide complete data 12 Months. Matching perfectly on sex, we obtain sampling weights, based on inverse of the mahalanobis distances for age, BMI, average step count at baseline and average weartime at baseline of patient $i$ and the participants in the donor pool. We select $M$ donors using sampling weights. For each $M$, we randomly select 5 weekdays with replacement and 2 weekend days with replacement to impute the missing week. 

After all intervals in all non-empty intervals $\mathcal{I}_i$ are imputed, $M$ complete epoch-level datasets are formed. Analyses can be implemented on each of these complete datasets, and the results combined with  mean daily step count and variance from $M$ datasets with Rubin's rules.

\begin{figure}
\center
\includegraphics[scale=0.5]{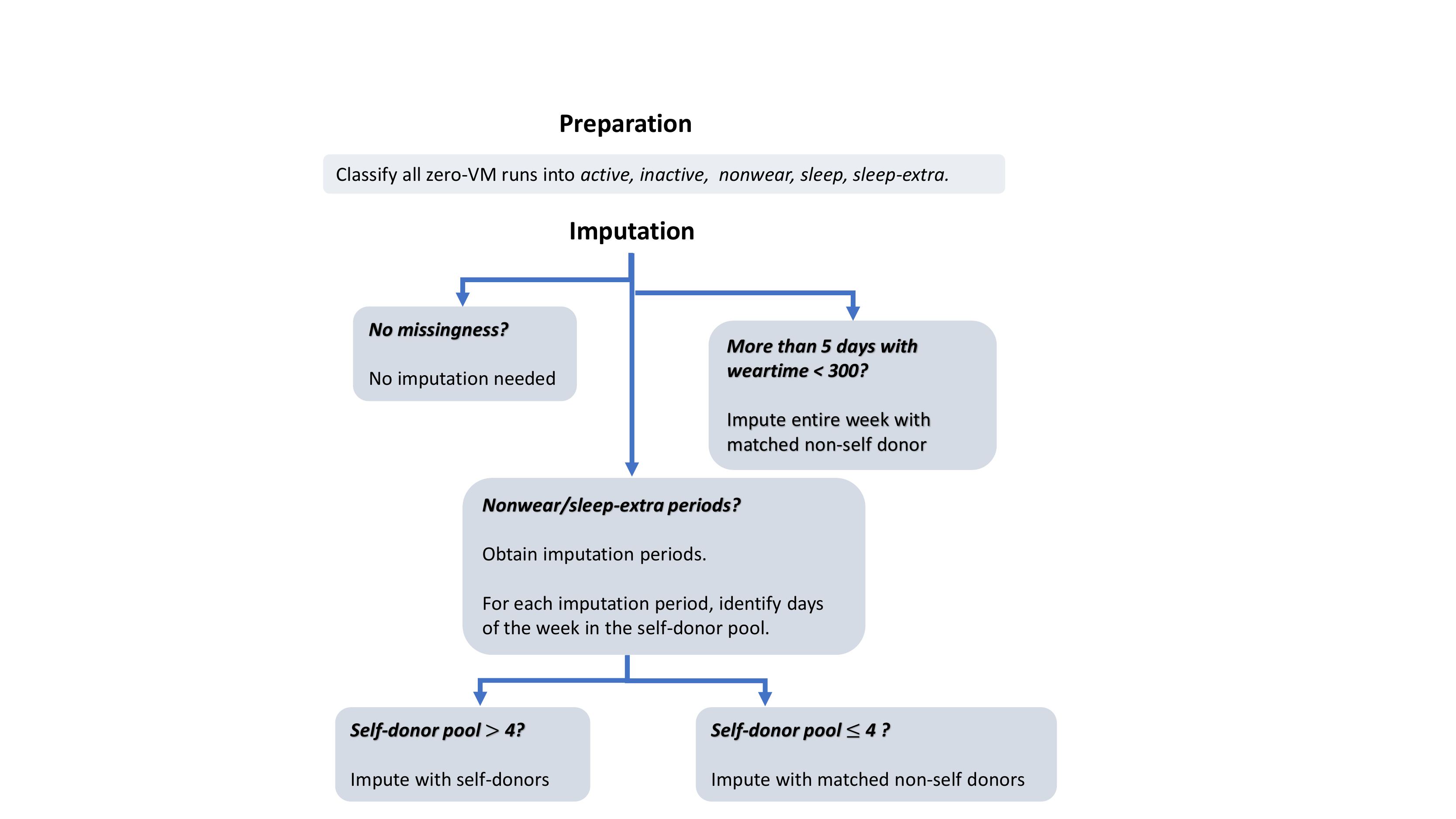}
\caption{Schematic for donor-based approach}
\label{Imputation schematic}
\end{figure}

A simplified schematic is shown Figure \ref{Imputation schematic}. \texttt{R} Code is for the non-parametric approach and an associated Vignette is provided in Supplementary Materials. 

\section{Simulation Studies}
\label{Sim}
We conducted simulation studies to establish statistical properties of the proposed approaches to handling missing data at the epoch-level. In the first simulation, we consider the setting where there is data from one time point - at 12 Months only. We assess the performance of the approaches in estimating the mean and standard error of the average step count across the week. In the second simulation, we consider the setting where data is collected at two time points, baseline and 12 Months. We assess the performance of the approaches in estimating the coefficients of a linear model where the average step count at 12 Months is regressed on average baseline step count and treatment arm. We also explore the estimates of the correlation between the average step count at baseline and the average step count at 12 Months.

\subsection{Simulation for One Time Point}

The first simulation aims to compare the statistical properties of parametric and non-parametric methods in estimating the mean and standard error of average weekly step count at one time point.

Step count data on the epoch-level have complex distributions that are difficult to characterize using a parametric model; devising a data generating model which adequately captures this complexity is difficult. Therefore, in our simulations, we take the approach of using bootstrap resamples of data from the PACE-UP trial from patients who have complete data, inducing missingness in the data, and comparing different approaches to handling the missingness. There were 438 patients at 12 Months who have completely observed data in the PACE-UP trial (150 in the control group, 150 in the postal group and 138 in the nurse group). In each repetition of the simulation study, we obtain a bootstrap sample where 120 patients in each treatment group is sampled without replacement. This creates a sample of 360 patients with perfect data. Ensuring that we sample without replacement is important, since having exact copies of patients in a dataset would put the non-parametric approach at an advantage since it uses donor pools from other patients.  

We then generate missingness under the following two scenarios: 
\begin{itemize}
\item Scenario 1: For a randomly selected $45\%$ of these 360 patients, sleep-extra periods and/or nonwear periods are induced in the following way:  we randomly select a patient from the PACE-UP trial who has incomplete data at 12 Months, and induce sleep-extra and/or nonwear periods according to the randomly selected patient's missingness pattern. This ensures that missingness is generated in a way that is representative of what is observed in a real life setting.
\item Scenario 2: In addition to inducing sleep-extra and/or nonwear periods in  $45\%$ of patients, a randomly select $2\%$ of patients provide no data for the entire week. 
\end{itemize}

We then consider the following methods of handling the incomplete data: 

\begin{itemize}
\item \textit{Available Case}: As a benchmark, we analyse the data as if it were the observed data, making no attempt to handle the missingness;
\item \textit{Complete Case}: Participants who provide at least 1 day of at least 540 minutes of weartime at 12 Months are included. The daily step count for any day with less than 540 minutes is set to missing. The average of the non-missing days are computed at baseline and at 12 months. 
\item \textit{Non-parametric Multiple Imputation}, as described in  \nameref{Non-para approach}, where age, sex and BMI are used as matching variables to sample non-self donors. We set $M=10$;
\item \textit{Parametric Multiple Imputation}, with specific and generic upper bounds, as described in  \nameref{Para approach}. We include BMI, sex and age as covariates in the imputation model, and set $M=10$. 
\end{itemize}

The estimands of interest are the mean and standard error of the average weekly step count at 12 months. We fit the following regression model for each arm separately (control, postal, and nurse):

\begin{equation} 
\bar{y}_{i, 1, .,.} = \beta_0 +\epsilon_{i},
\end{equation}

where we assume that $e_{i} \sim N(0, \sigma^2)$, and obtain the estimate and standard error of $\beta_0$. 

The approaches to handling missing data are assessed by comparing bias in the estimate of the mean, and increase in standard error compared to the true value. 

We run 100 repetitions. Multiple Imputation using Tobit regression is conducted in \texttt{STATA} and all other aspects of the simulation are conducted in \texttt{R}. 

\subsubsection{Results}

\begin{figure}[H]
\center
\includegraphics[scale=0.6]{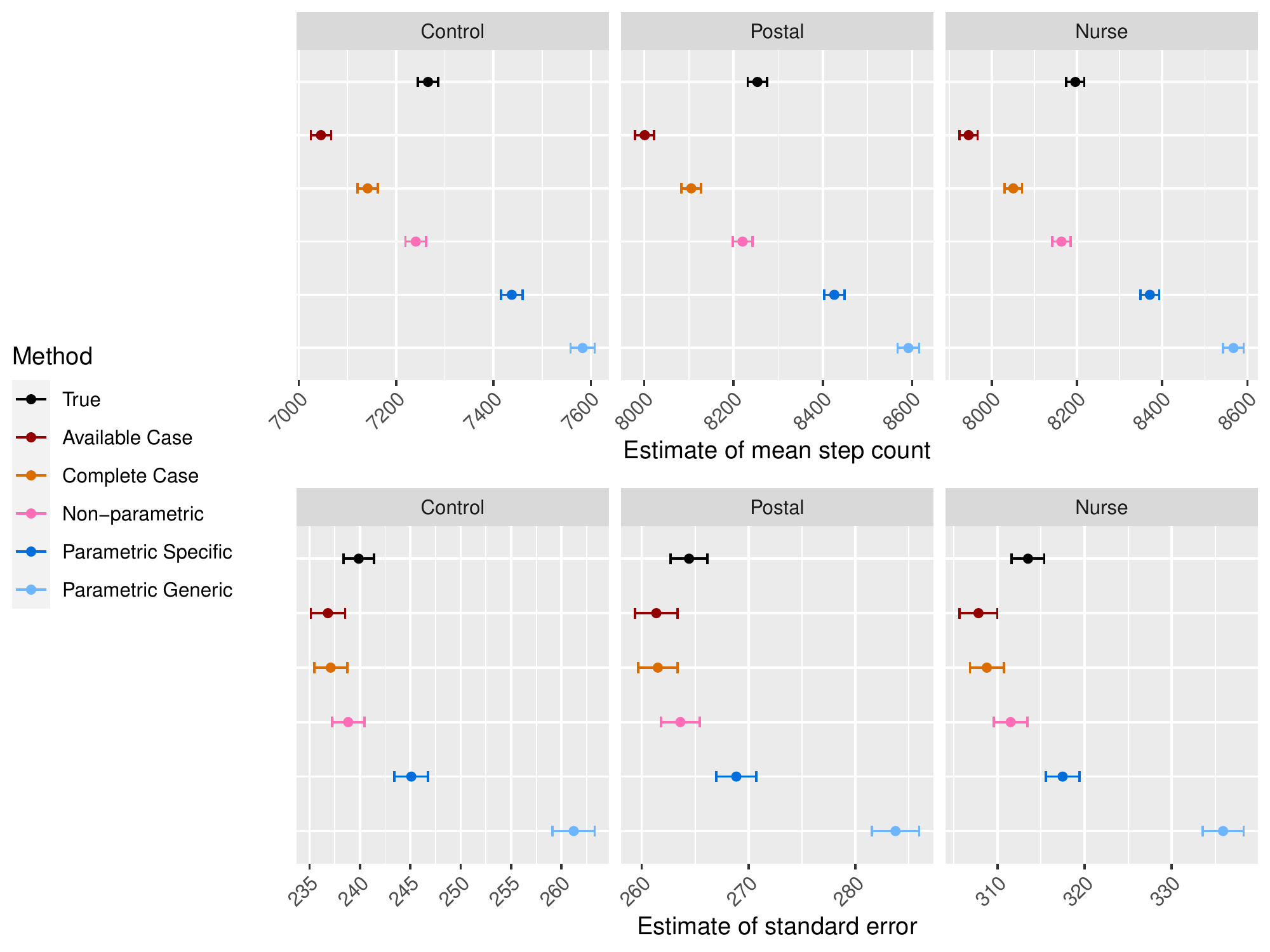}
\caption{Results for Simulation for one time point: Scenario 1. Results are shown by arm. For each method, estimates for the mean step count are shown in the left panel, and estimates for the standard error of the mean, are shown on the right panel. The error bars indicate $\pm 1.96 \times$ MC error.}
\label{sim_full_12M_prop0.45_norep_mixw}
\end{figure}

Results of the simulation with Scenario 1, where $45\%$ of participants have nonwear and/or sleep extra, are shown in Figure \ref{sim_full_12M_prop0.45_norep_mixw}. The top panels display the estimates of the mean within each arm. We observe that available case leads to a downward bias of over 200 steps in all arms. This is expected as the available case assumes that the data with induced missingness is complete. Complete case analysis leads to a slight downward bias. The non-parametric approach leads to estimates that are closest to the true value; they are within MC error, but it appears that there is a small downward bias. Both parametric approaches - with the generic and specific upper bound - lead to upward bias, with the generic upper bound in particular leading to an upward bias of over 300 steps in all arms. 

The bottom panels of  Figure \ref{sim_full_12M_prop0.45_norep_mixw} shows the estimate of the standard error. The available case approach leads to a slight decrease in SE which is likely due to the fact that the dataset with missingness nonwear and/or sleep extra has a lower mean, and its standard error is generally lower. The complete case analysis also leads to a slight decrease in SE. For the non-parametric approach, we observe that the standard error is within the MC error of the true standard error, but appears to be slightly smaller. This is due to the slight downward bias in the estimates of the mean in the non-parametric approach. Finally, we observe that the parametric approaches lead to comparatively larger increases in SE, particularly when a generic upper bound is used. 

\begin{figure}[H]
\center
\includegraphics[scale=0.7]{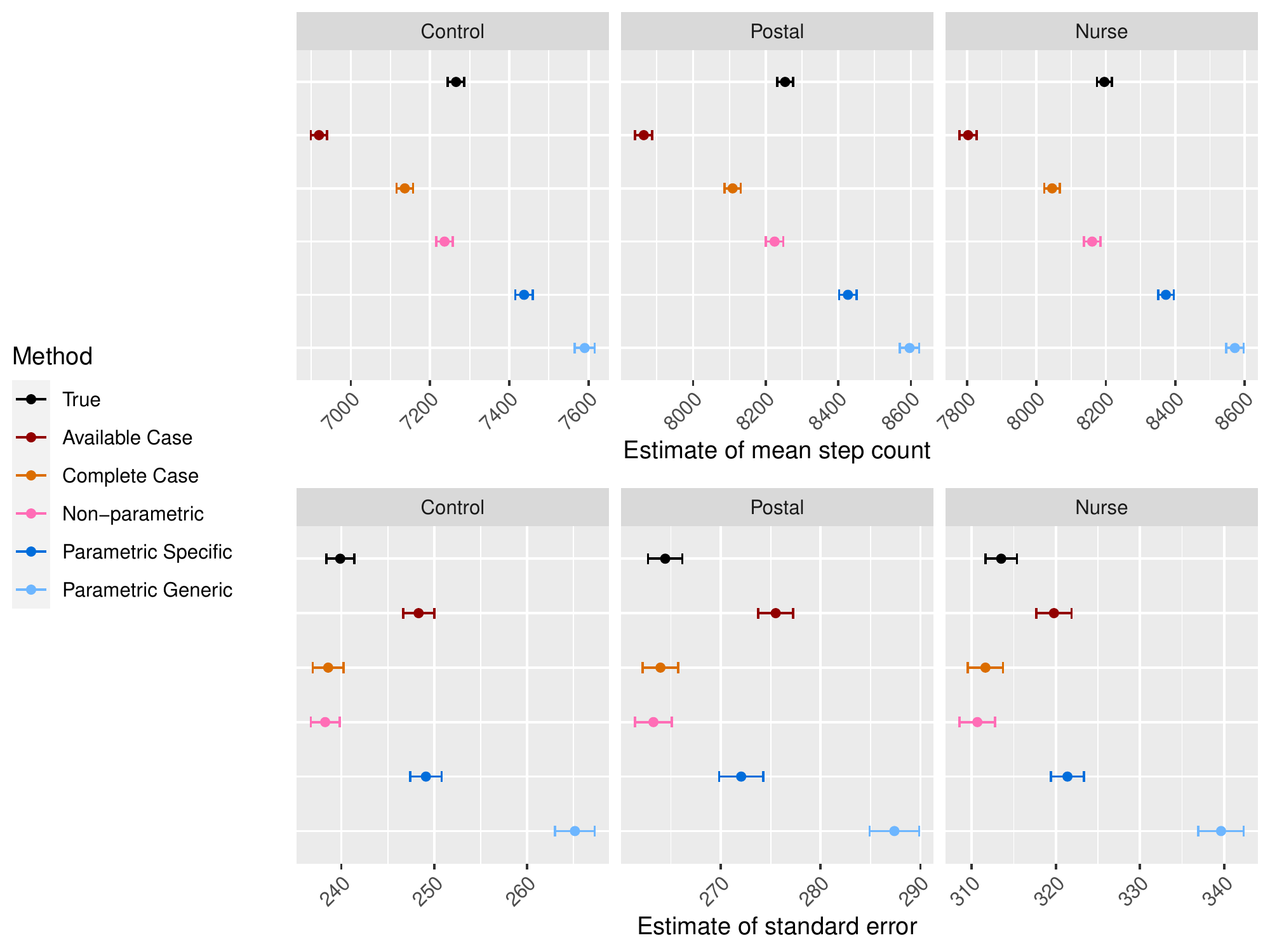}
\caption{Results for Simulation for one time point: Scenario 2. Results are shown by arm. For each method, estimates for the mean step count are shown in the left panel, and estimates for the standard error of the mean, are shown on the right panel. The error bars indicate $\pm 1.96 \times$ MC error.}
\label{sim_full_12M_prop0.45_ww0.02_norep_mixw}
\end{figure}

Results of the simulation with Scenario 2, where $45\%$ of participants have nonwear and/or sleep extra and an additional $2\%$ of patients have no data for the entire week, are shown in Figure \ref{sim_full_12M_prop0.45_ww0.02_norep_mixw}. The conclusions are similar to those given for Scenario 1, except for one difference. When there are entire weeks that have no data, the available case leads to estimates of the standard error that are much larger than in Scenario 1, since there the $2\%$ of patients with no data lead to much greater variability. 

In the case where there is data from one time point, we observe that the non-parametric approach to MI leads to correlations between baseline and 12M that are closest to the true value, and the point estimates are closest to the true value. 

\subsection{Simulation 2: Two Time Points}
\label{Sim2}
The second simulation explores the setting where there is data at baseline and 12 months, and there is missingness at 12 months. The aim is to assess the MI approaches in estimating the regression coefficients of a model which regresses the average step count at 12 months on the average step count at baseline and treatment arm. We also compare the MI approaches in estimating the correlation between the average weekly step count between baseline and 12 months within each arm. 
 
In this simulation, we obtain data from the 277 patients who have complete data at both baseline and 12 months in the PACE-UP trial  (95 in the control group, 92 in the postal group and 90 in the nurse group). Similarly to the previous simulation, in each repetition of the simulation study, we obtain a bootstrap sample where 85 patients in each treatment group is sampled without replacement. This creates a sample of 255 patients with perfect data for each repetition of the simulation. 

We then generate missingness under the following two scenarios: 
\begin{itemize} 
\item Scenario 1: for a randomly selected $45\%$ of these 255 patients, sleep-extra periods and/or nonwear periods are induced in the 12-Month data only by randomly selecting a patient from the PACE-UP trial who has incomplete data at 12 Months, and induce sleep-extra and/or nonwear periods according to their missingness pattern. 
\item Scenario 2: In addition to inducing sleep-extra and/or nonwear periods in  $45\%$ of patients, a randomly select $2\%$ of patients provide no data at 12 Months. 
\end{itemize}

The following approaches are used to handle the missing data:

\begin{itemize}
\item \textit{Complete Case}: Participants who provide at least 1 day of at least 540 minutes of weartime at 12 Months are included. The daily step count for any day with less than 540 minutes is set to missing. The average of the non-missing days are computed at baseline and at 12 months. 
\item \textit{Non-parametric Multiple Imputation}, as described in  \nameref{Non-para approach}. We BMI, sex, age, average step count at baseline and average weartime at baseline as matching variables where a non-self donor is needed. We set $M=10$;
\item \textit{Parametric Multiple Imputation}, with specific and generic upper bounds, as described in  \nameref{Para approach}. We include BMI, sex, age and average step count at baseline as covariates in the imputation model. We set $M=10$. 
\end{itemize}

The estimands of interest are the coefficients and standard errors of the following regression model:

\begin{equation} 
\label{sim2}
\bar{y}_{i, 1, ., .} = \beta_0 + \beta_1 \bar{y}_{i, 0, ., .} + \beta_2 \mathbb{I}(\mbox{arm}_i=\mbox{postal}) +\beta_3 \mathbb{I}(\mbox{arm}_i=\mbox{nurse}) + \epsilon_{i},
\end{equation}

where we assume that $e_{i} \sim N(0, \sigma^2)$. We assess bias in the estimates of the regression coefficients and increase in the standard error compared to the true value.  An additional estimand is the correlation of average step count across the week between Baseline and 12 months for each arm. Although Equation \eqref{sim2} assumes that these correlations are equal across arms, we wish to examine how well the approaches to missing data are able to preserve the correlation within each arm, as correlation is expected to be attenuated. 

We run 100 repetitions. Multiple Imputation using Tobit regression is conducted in \texttt{STATA} and all other aspects of the simulation are conducted in \texttt{R}.

\subsubsection{Results}
\begin{figure}[H]
\center
\includegraphics[scale=0.7]{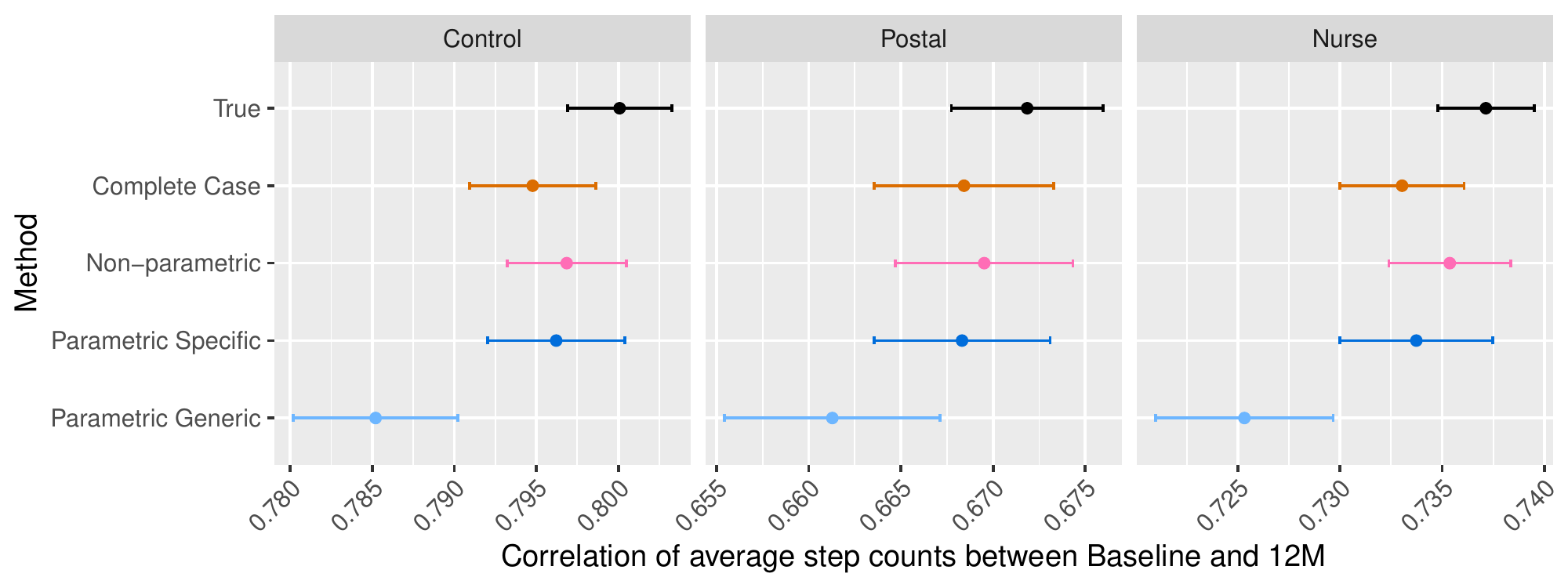}
\caption{Results for Simulation for two time points: Scenario 1. Results are shown by arm. For each method, estimates for correlation between baseline and 12 month average step count is displayed for $M=1$. The error bars indicate $\pm 1.96 \times$ MC error.}
\label{sim_l_12M_corr_prop045_boot_norep_ww_mixw}
\end{figure}

\begin{figure}[H]
\center
\includegraphics[scale=0.7]{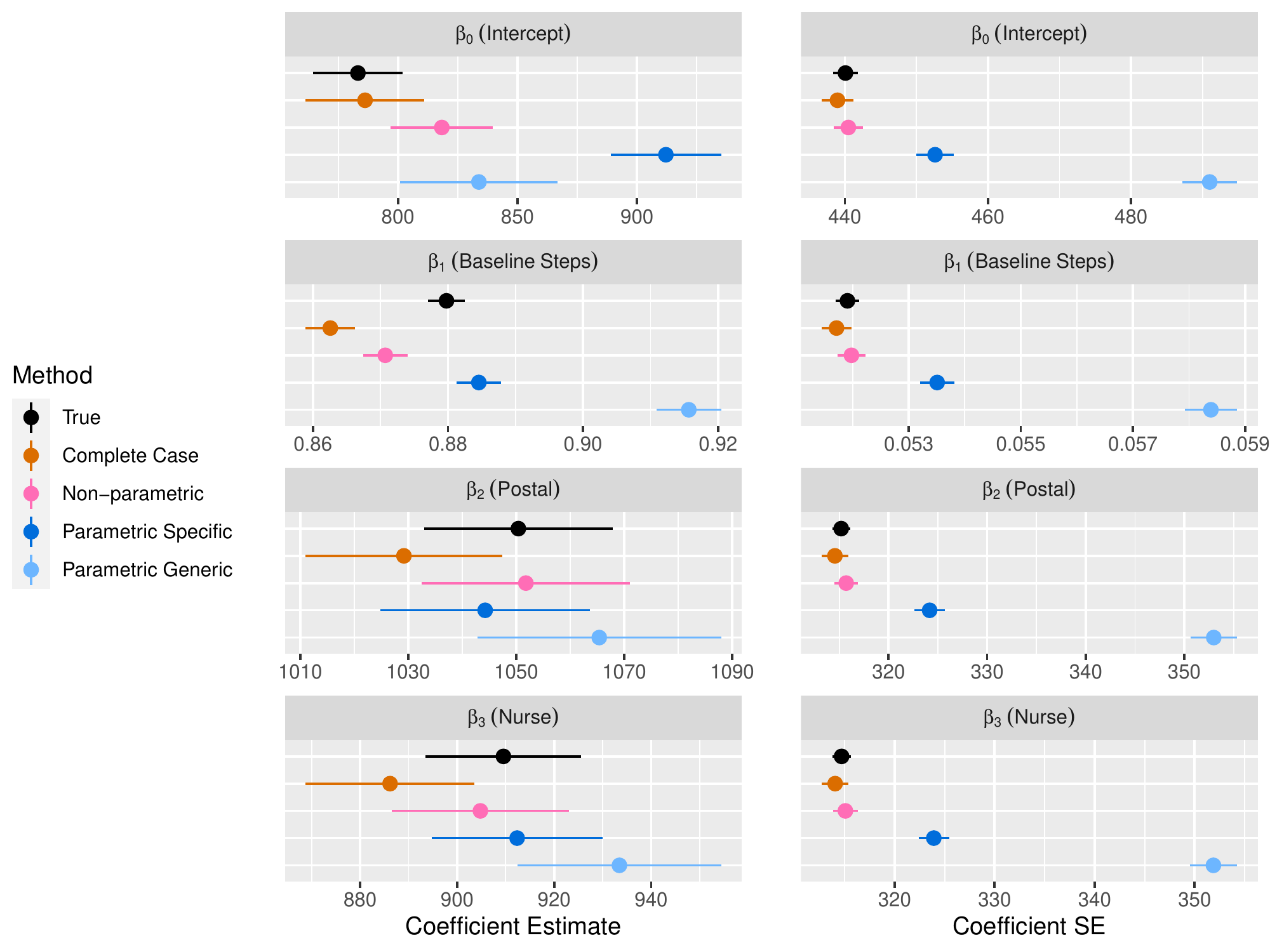}
\caption{Results for Simulation for two time points: Scenario 1. Results are shown by arm. For each method, regression coefficients and standard errors for Equation \eqref{sim2} are displayed. The error bars indicate $\pm 1.96 \times$ MC error.}
\label{sim_l_12M_baseadj_prop045_boot_norep_ww_mixw}
\end{figure}

Results for simulation with Scenario 1, where $45\%$ of participants have nonwear and/or sleep-extra, are displayed in Figures \ref{sim_l_12M_corr_prop045_boot_norep_ww_mixw} and \ref{sim_l_12M_baseadj_prop045_boot_norep_ww_mixw}. In Figure \ref{sim_l_12M_corr_prop045_boot_norep_ww_mixw}, the correlation between the average step count between Baseline and 12 Months are displayed for each arm. We observe that the true correlation is lower in the treatment groups compared to the control group. Since the treatments were both effective in increasing participants' step count at 12 months, it is unsurprising that the correlation between baseline and 12 months is lower for the treatment groups. We observe that the correlation is attenuated the least for the non-parametric approach. Using generic upper bound leads to greater attenuation than the specific upper bound when using a parametric approach. The correlations for the complete case analysis appears to be comparable to that of the parametric approach with specific upper bound.

In Figure \ref{sim_l_12M_baseadj_prop045_boot_norep_ww_mixw}, results for estimates of the mean and standard error of the effects of the regression model in Equation \eqref{sim2} are shown for Scenario 1. The estimates for the intercept produced by the non-parametric and complete case analysis are within MC error of the true values; the non-parametric approaches lead to slightly higher estimates. The parametric approaches result in an upward bias of the intercept. For the effect of average stepcount at baseline, the parametric approach with specific upper bound leads to the least biased estimate. The non-parametric approach and complete case approach are downward biased, and the parametric approach with generic upper bound leads to a large upward bias. Of particular interest are the effects of treatment (postal and nurse), which are both estimated well by the non-parametric approach and the parametric approach with specific upper bound. The complete case approach leads to a slightly lower estimate, and the parametric approach with generic upper bound leads to a slightly larger value of estimated effect. Across all coefficients, we observe that the non-parametric approach and complete case analysis lead to the smallest standard errors, and the parametric approach with person-specific upper bounds leads to smaller standard errors than the generic upper bounds. Overall, we observe that the non-parametric and parametric approach with person-specific upper bounds provide estimates of the treatment effect that are least biased, and the non-parametric approach is additionally provides a smaller standard error.

\begin{figure}[H]
\center
\includegraphics[scale=0.7]{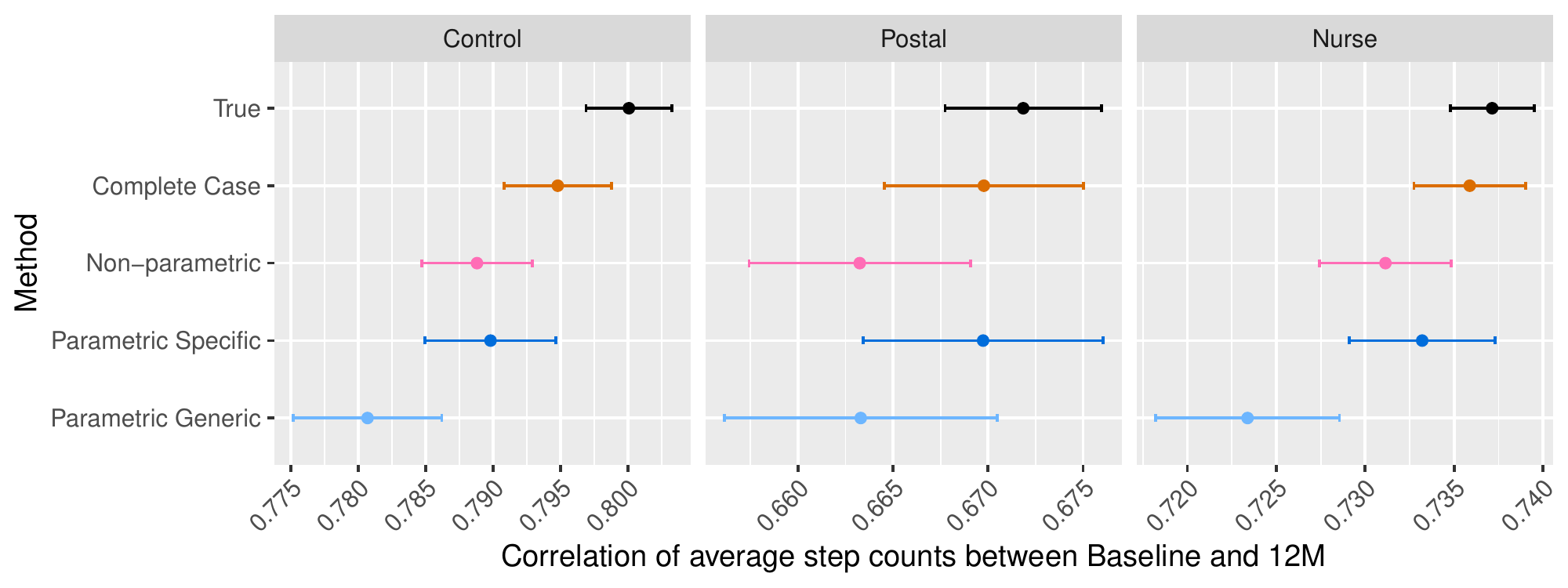}
\caption{Results for Simulation for two time points: Scenario 2. Results are shown by arm. For each method, estimates for correlation between baseline and 12 month average step count is displayed for $M=1$. The error bars indicate $\pm 1.96 \times$ MC error.}
\label{sim_l_12M_corr_prop045_ww002_boot_norep_ww_mixw}
\end{figure}

\begin{figure}[H]
\center
\includegraphics[scale=0.7]{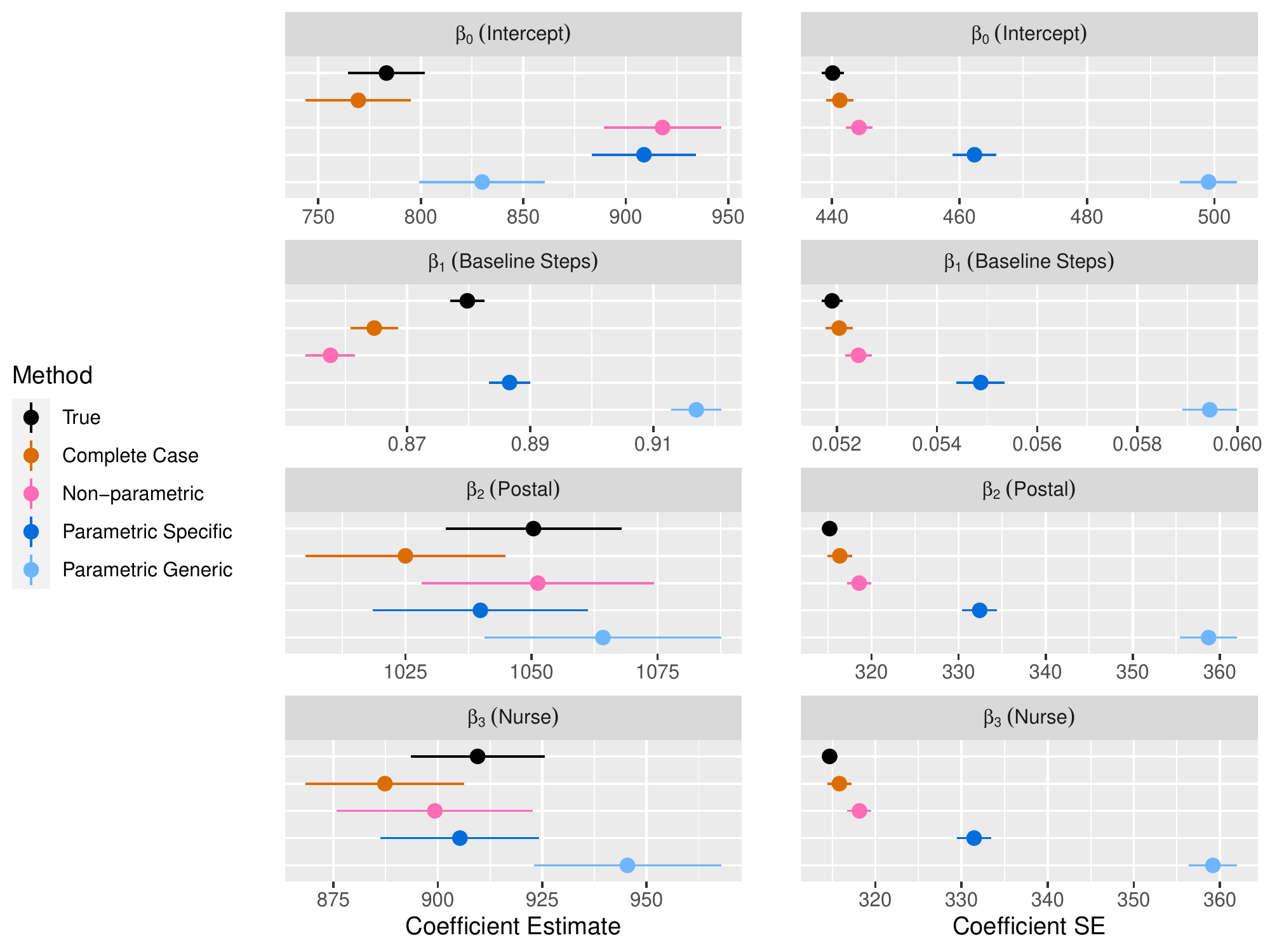}
\caption{Results for Simulation for two time points: Scenario 2. Results are shown by arm. For each method, regression coefficients and standard errors for Equation \eqref{sim2} are displayed. The error bars indicate $\pm 1.96 \times$ MC error.}
\label{sim_l_12M_baseadj_prop045_ww002_boot_norep_ww_mixw}
\end{figure}

Results for simulation with Scenario 2, where $45\%$ of participants have nonwear and/or sleep-extra at 12 Months, and an additional $2\%$ of patients have no data at at 12 Months, are displayed in Figures \ref{sim_l_12M_corr_prop045_ww002_boot_norep_ww_mixw} and \ref{sim_l_12M_baseadj_prop045_ww002_boot_norep_ww_mixw}. In Figure \ref{sim_l_12M_corr_prop045_ww002_boot_norep_ww_mixw}, the correlation between the average step count between Baseline and 12 Months are displayed for each arm. We observe that the correlation for the non-parametric approach and parametric approach with specific upper bound have a similar level of attenuation for the control and nurse groups; the non-parametric approach leads to slightly more attenuation in the postal group. The parametric approach with the generic approach leads to a greater amount of attenuation compared to the other imputation approaches. The complete case analysis appears to retain the correlation better than other approaches in this setting. Compared to Scenario 1, the non-parametric approach appears slightly less effective at preserving the correlation between baseline and 12 months when there are participants with the entire week missing.   

In Figure \ref{sim_l_12M_baseadj_prop045_ww002_boot_norep_ww_mixw}, results for estimates of the mean and standard error of the effects of the regression model in Equation \eqref{sim2} are shown for Scenario 2. All approaches except the complete case analysis lead to upward bias in the estimate of the intercept; the non-parametric approach and parametric approach with specific upper bound lead to a similar amount of bias.  For the effect of average stepcount at baseline, the parametric approach with specific upper bound leads to estimates that are closest to the true values. The non-parametric approach and complete case analysis lead to a downward bias, while the parametric approach with generic upper bound leads to a large upward bias. The effects of treatment (postal and nurse) are estimated with least bias for the non-parametric approach. Complete case analysis leads to a downward bias, while the  parametric approach with generic upper bound leads to slightly larger values of estimated effect. We again observe that the that the non-parametric approach and complete case analysis lead to the smallest standard errors, and the parametric approach with person-specific upper bounds leads to smaller standard errors than the generic upper bounds. Thus, focusing specifically on the effects of treatment, we observe that the non-parametric approach leads to estimates that are least biased and with smallest standard error.

In the simulations for one time point, we  observed a very strong advantage for the non-parametric approach compared to other approaches. When there is data from two time points, we still observe that the treatment effects are estimated best by the non-parametric approach; the estimates are least biased, and standard errors are much smaller than those produced by the parametric approach. For the estimates of the intercept and effect of the average baseline stepcount, there is no clear advantage of the non-parametric approach, particularly when whole-week imputation is required.
Due to the relatively small sample size used in the simulation, we note that the pool of non-self donors becomes limited to a small number of patients for the non-parametric approach, which may hinder slightly its performance in this setting.

\section{Application to the PACE-UP trial}
\label{App}
We now apply the proposed approaches to handling missingness to an analysis of the PACE-UP trial. In the previous section, missingness was induced in the simulated datasets; we now demonstrate the applicability of the proposed approaches to a genuine trial dataset with real instances of missingness. Table \ref{paceup_missing} illustrates the breakdown of the total 1023 participants into the types of missingness observed. We classify participants by whether they have completely observed data, non-wear periods only, sleep-extra only, or both non-wear and sleep-extra. Participants may also have more than 5 days with weartime $<540$ minutes, in which case the entire week is imputed in the non-parametric approach. These participants are excluded in the complete case approach. We also illustrate the proportion of participants that have zero days with wear time $<540$ minutes, between 1 and 5 days (inclusive), and greater than 5 days. There is a greater amount of missingness at 12 months compared to baseline.

\begin{table}[H]
\centering
\begin{tabular}{|r|rr|}
\hline
\multicolumn{1}{|l|}{\textbf{}}                                                    & \multicolumn{1}{c}{\multirow{2}{*}{\textbf{Baseline}}} & \multicolumn{1}{c|}{\multirow{2}{*}{\textbf{12 Months}}} \\ \cline{1-1}
\multicolumn{1}{|l|}{\textbf{Type of missingness}}                                 & \multicolumn{1}{c}{}                                   & \multicolumn{1}{c|}{}                                    \\ \hline
Completely observed                                                                & 554 (54.2\%)                                           & 473 (46.2\%)                                             \\
Non-wear only                                                                      & 361 (35.3\%)                                           & 287 (28.1\%)                                             \\
Sleep-extra only                                                                   & 51 (4.99\%)                                            & 104 (10.2\%)                                             \\
Non-wear and sleep-extra                                                           & 57 (5.57\%)                                            & 94 (9.19\%)                                              \\
Whole week imputation                                                              & 0 (0\%)                                                & 65 (6.35\%)                                              \\ \hline
\multicolumn{1}{|l|}{\textbf{Number of days with weartime \textless{}540 minutes}} & \multicolumn{2}{r|}{}                                                                                             \\ \hline
0                                                                                  & 696 (68.0\%)                                           & 562 (54.9\%)                                             \\
Between 1 and 5                                                                    & 327 (32.0\%)                                           & 396 (38.7\%)                                             \\
Greater than 5                                                                     & 0 (0\%)                                                & 65 (6.35\%)                                              \\ \hline
\multicolumn{1}{|l|}{Total}                                                        & \multicolumn{2}{c|}{1023}                                                                                         \\ \hline
\end{tabular}
\caption{The number of patients with each of the missing types and their percentages are shown for PACE-UP trial data at Baseline and 12 Months. }
\label{paceup_missing}
\end{table}

We analyse the data using a linear model which regresses the average step count at 12 months on the average step count at baseline, arm, and primary care practice: 

\begin{align}
\label{paceup_model}
\begin{split}
\bar{y}_{i,2,., .} &= \beta_0 + \beta_1 \bar{y}_{i,0,., .} + \beta_2 I\left( \mbox{arm}_i= \mbox{postal} \right) \\
&+\beta_3 I\left( \mbox{arm}_i=\mbox{nurse} \right) + \beta_4 P2_i + \beta_5 P3_i + .. + \beta_9 P7_i + \epsilon_{i}, 
\end{split}
\end{align}

where $P2, P3, ..., P7$ are dummy variables for the primary care practice that the participant resides in. We assume that $e_{i} \sim N(0, \sigma^2)$. \\

We note that this is a simpler analysis compared to the primary analysis of the PACE-UP trial, where additional covariates (sex and age group) were included, and a clustering effect was included to account for household, since a small number of participants were in couples. For the purposes of comparing the approaches to handling missing data, we use a simpler model to analyse the results. \\

The following methods of handling missing data are considered: 
\begin{itemize}
\item \textit{Complete Case}: Participants who provide at least 1 day of at least 540 minutes of weartime at 12 Months are included. The daily step count for any day with less than 540 minutes is set to missing. The average of the non-missing days are computed at baseline and at 12 months. We use the weartime calculated by the Actilife software. 
\item \textit{Non-parametric Multiple Imputation}, as described in  \nameref{Non-para approach}. Imputation is conducted firstly in the baseline dataset, separately in the three treatment arms. We use BMI, sex and age as matching variables where a non-self donor is needed. The average of the imputed baseline average stepcounts are computed, which is then used for the imputation at 12 Months, which is again conducted separately in the three treatment arms. At 12 Months, we use BMI, sex, age, average step count at baseline and average weartime at baseline as matching variables where a non-self donor is needed. 
\item \textit{Parametric Multiple Imputation}, with specific and generic upper bounds, as described in  \nameref{Para approach}. The seven days at baseline, and seven days at 12 Months are modelled as jointly normally distributed, conditional on covariates BMI, sex, age. Imputation is performed separately in each arm. 
\end{itemize}

\begin{figure}[H]
\center
\includegraphics[scale=0.7]{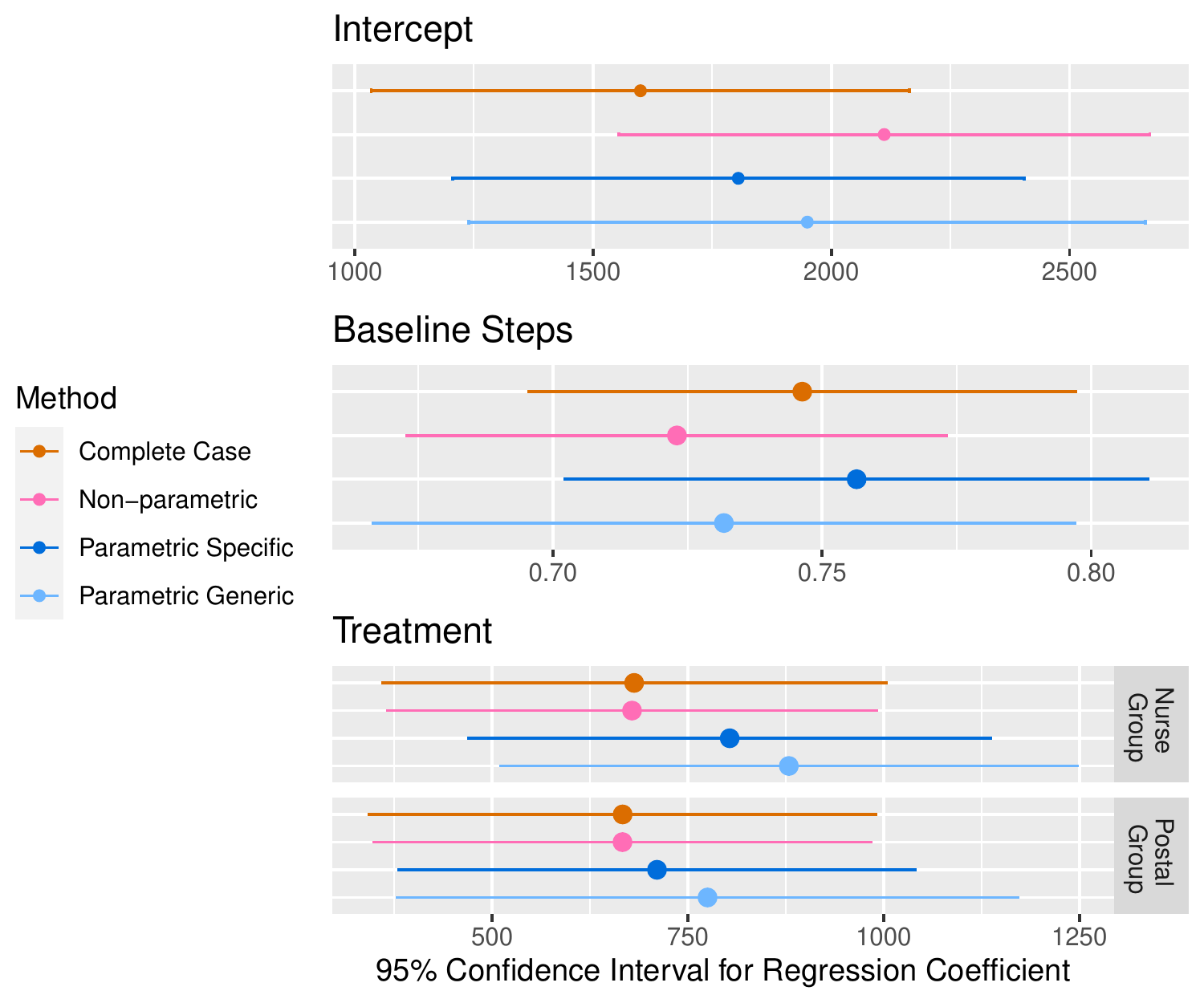}
\caption{Results for the analysis of the PACE-UP trial. For each method of handling missing data, the $95\%$ confidence intervals for the effects for the regression model in Equation \ref{paceup_model} is displayed. The $95\%$ confidence intervals for the effects of practices are shown in Figure \ref{result_practice} in the Appendix. } 
\label{result}
\end{figure}

Results for the estimated effects effects and their estimated confidence intervals produced by each method are displayed in Figure \ref{result}. The means and standard errors of the effects are displayed in Table \ref{results_tab}. The confidence intervals across the methods overlap, but we observe noticeable differences in the point estimates and standard errors which reveal the potential impact that missing data assumptions can have on the results of the primary analysis. \\

\begin{table}[H]
\begin{tabular}{|l|llll|}
\hline
\multirow{2}{*}{Coefficient} & \multicolumn{4}{c|}{\textbf{Estimated Means and Standard Errors}}              \\ \cline{2-5} 
                             & Complete Case    & Non-parametric   & Parametric Specific & Parametric Generic \\ \hline
Intercept & 1599.65 (288.05) & 2110.96 (284.29) & 1804.87 (305.98) & 1949.72 (362.10) \\ 
Baseline Steps &    0.75 (  0.03) &    0.72 (  0.03) &    0.76 (  0.03) &    0.73 (  0.03) \\ 
Postal 
 Group &  666.69 (166.00) &  666.49 (163.12) &  710.55 (169.37) &  775.10 (203.06) \\ 
Nurse 
 Group &  681.35 (165.05) &  678.65 (160.16) &  803.41 (170.95) &  878.96 (188.94) \\ 
P2 & -831.83 (266.32) & -658.93 (255.24) & -581.28 (282.59) & -422.90 (321.30) \\ 
P3 & -287.51 (256.00) & -446.34 (248.35) & -395.47 (269.78) & -312.76 (293.50) \\ 
P4 & -658.03 (280.20) & -511.25 (269.67) & -369.16 (293.85) & -348.26 (327.41) \\ 
P5 &   35.25 (248.72) &  184.13 (240.16) &  194.49 (258.25) &  337.40 (299.67) \\ 
P6 &   78.29 (265.13) &  136.54 (257.50) &  165.02 (291.26) &  325.56 (316.06) \\ 
P7 &  -65.11 (332.66) &  110.05 (324.23) &  -18.34 (341.61) &   12.28 (379.22) \\ \hline
\end{tabular}
\caption{Results for the analysis of the PACE-UP trial. For each method of handling missing data, the estimated means (with standard errors in parentheses) are shown for the effects for the regression model in Equation \ref{paceup_model} is displayed. Note that the Practices have been included to reflect the design of the study, but their coefficients should not be the focus of the interpretation.}
\label{results_tab}
\end{table}

In the point estimates, we find patterns that are similar to what we observed in \nameref{Sim2}. For the intercept, the complete case approach leads to the smallest estimate, and the non-parametric approach leads to the highest estimate. When comparing the coefficients for the average baseline stepcount, it should be noted that the values of the average baseline stepcounts are computed using the imputed values at baseline for the non-parametric and parametric approaches, so their values are different across approaches. In the effects for treatment, we observe that the complete case and non-parametric approach leads to the smallest point estimates, and the parametric approaches lead to the largest estimates. \\

Across all coefficients, we find that the standard errors are smallest for the non-parametric and complete case approaches; the non-parametric approach leads to slightly smaller standard errors. The parametric approaches produce larger standard errors; the generic upper bound, in particular, leads to the largest standard errors. These observations are consistent with what we found in the simulation studies. \\

\begin{figure}[H]
\center
\includegraphics[scale=0.7]{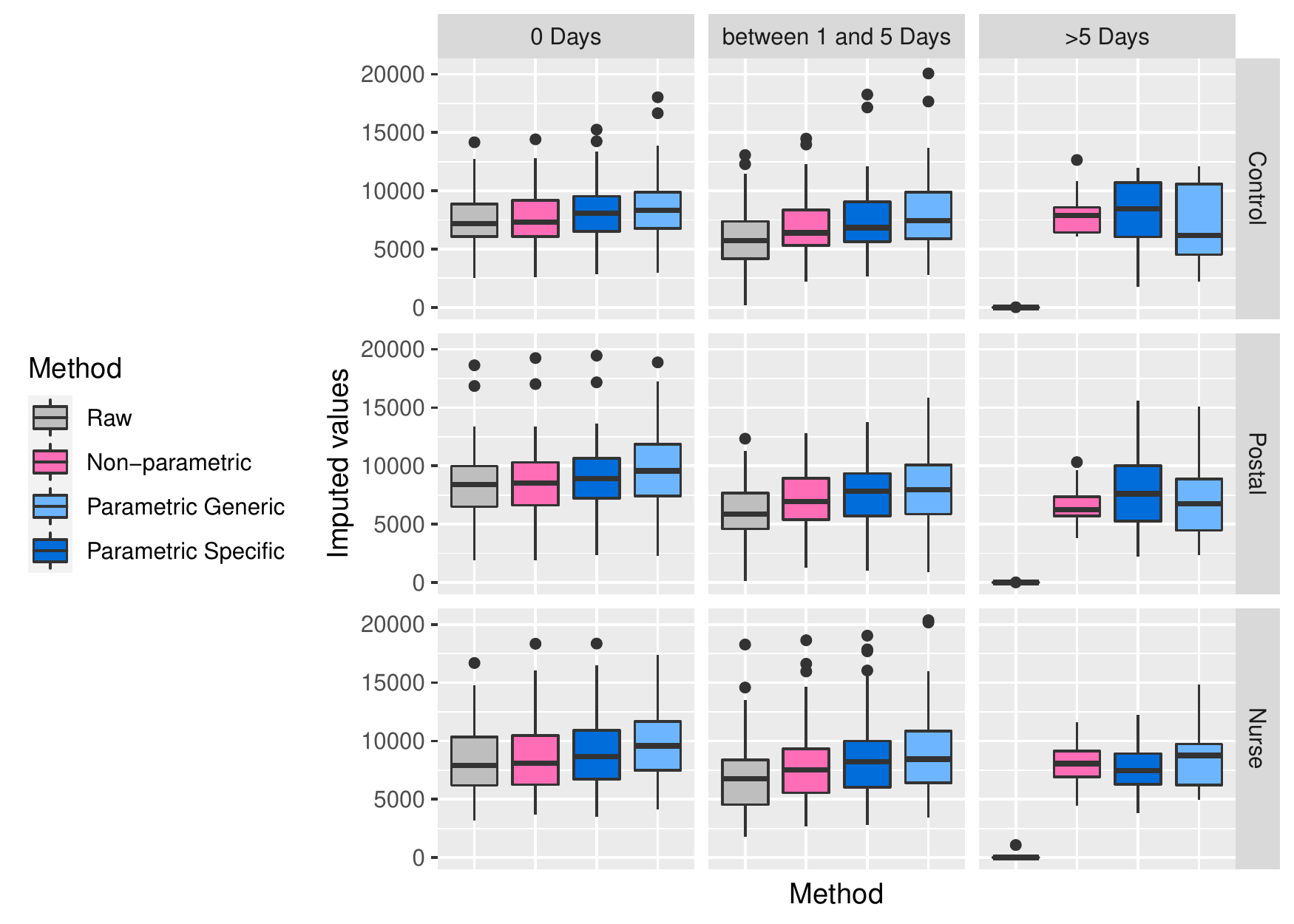}
\caption{Boxplots compare the raw value with the imputed values at 12 Months under the non-parametric and parametric approaches when there are zero days with weartime $< 540$ minutes (left panel) and between 1 and 5 days with weartime $< 540$ minutes (middle panel) and more than 5 days with weartime $< 540$ minutes (right panel). Results are shown by treatment group.}
\label{Year1_boxplots_bygroup}
\end{figure}

Figure \ref{baseline_boxplots_bygroup} in the Appendix displays boxplots comparing the raw values with the imputed values for $M=1$ for each treatment group at baseline. We display the boxplots separately by the extent of missingness, classifying participating as having zero days with weartime $< 540$ minutes or between 1 and 5 days with weartime $<540$ minutes.  We observe that the non-parametric approach leads to a slightly larger values compared to the raw values, the parametric approach with specific upper bound leads to slightly higher values than that. The parametric approach with generic upper bound, which does not use the epoch-level information on missingness, leads to the highest imputed values. In Figure \ref{Year1_boxplots_bygroup}, we observe the equivalent boxplots for 12 Months, where there is an additional column for participants where the entire week is missing. These participants are excluded in the complete case analysis. In this column, we find that distributions have more skewness. In particular, the non-parametric approach, which imputes the whole week with non-self donors, leads to very small variability in imputed values compared to the parametric approaches. \\ 

The original analysis of the PACE-UP trial reported an estimated effect of the postal intervention of 642 (with standard error 160) and effect of the nurse intervention of 677 (with standard error of 159) when using a complete case analysis \citep{Harris2017}. Their imputation analysis which includes participants with missing data at 12 Months uses a day-level imputation model with the following variables: treatment group, baseline steps, gender, age, practice, and month of baseline accelerometry. This analysis produced an estimated effect of the postal intervention of 638 (with standard error 160) and effect of the nurse intervention of 679 (with standard error of 159). While a precise comparison with our re-analysis cannot be made since the analysis models differ, we note that the estimates for the effects of treatment produced by the epoch-level non-parametric approach are most similar to those produced in the original study. \\

The purpose of this analysis is to compare the impact of the differing approaches to missing data on the model results. We have therefore kept the analysis and imputation models relatively simple, and assume that the data are MAR given the observed data. A full analysis should include sensitivity analyses to the MAR assumption. This would involve careful consideration of appropriate departures from the MAR assumption. For example, one might consider the following scenarios: 

\begin{itemize} 
\item If participants take the accelerometer off too early in the evening, or put it on late in the morning, it may be because they are at home and not being particularly active. Under this assumption, we would impute values that are lower than that assumed under MAR. 
\item If participants did not provide sufficient wear time for more than 5 days a week (which would lead to whole-week imputation in the non-parametric approach), this may be because the participant is less active than usual during the week and does not feel motivated to wear the device. Under this assumption, we would impute values that are lower than that assumed under MAR.
\item Some participants may remove the device while they are exercising as they find it uncomfortable to wear. In this case, it is possible that values above that assumed under MAR should be imputed. However, identifying periods where participants removed the device for this reason is difficult to discern without information from, for example, activity reports from participants. 
\end{itemize}

Further, in addition to considering departures from the MAR assumption, a full analysis should take into account additional auxiliary variables in the imputation model, such as weather variables, which have shown to be predictive of daily step count and also of whether daily step count is missing \citep{Tackney2021}.

\section{Discussion}

This paper described the challenges of applying MI to epoch-level accelerometer data; namely, the difficulty in identifying and classifying missingness, and the complicated nature of epoch-level distributions. Possible methods of overcoming these challenges are presented. Firstly, a novel approach to classifying epoch-level zero-count periods into inactive, nonwear, sleep and sleep-extra periods is presented, which carefully teases out differences between per-protocol instances of no activity, short periods of inactivity which are not missing data, and periods where missing data is incurred due to participants removing the device. These missing periods can be handled with either parametric and non-parametric Multiple Imputation. In the parametric approach, step counts on the day-level are imputed using epoch-level information on the missing periods per day to specify an informed upper bound for Tobit regression. In the non-parametric approach, missing periods are replaced by self- or non-self donors. 

These approaches were compared using simulations where zero-count periods are generated using missingness patterns observed in the dataset. Simulation studies conducted in the literature to date often induce missingness in hour-long chunks, which does not reflect the complexities of how missingness arises in practice in the accelerometer setting. The simulations showed the merits of using the epoch-level information in MI. In the setting where the average step count for each treatment group is estimated for one time point, the non-parametric approach leads to estimates with the least bias and highest precision. The parametric approach with the specific upper bound leads to less bias and more precision than the parametric approach with the generic upper bound.  Where data has been collected over two time points and the analysis model regresses the average step count at 12 Months on the average step count at baseline and treatment group, the non-parametric approach and parametric approach with specific upper bound have a more comparable performance, but the non-parametric approach leads to least biased point estimates for the treatment effects while maintaining a small standard error. By considering settings where there is data for just one time point as well as two time points, these findings are relevant to cross-sectional as well as longitudinal analyses of accelerometer data.

We performed a re-analysis of the 2017 PACE-UP trial, where the results for a simplified analysis model using complete case, the non-parametric approach, and parametric approaches with specific and generic upper bounds are compared. While the approach to missingness does not overall change conclusions of the study, they point to potentially important implications for results. In particular, we observe that estimated effects of treatment are slightly higher for the parametric approaches, and the standard errors are larger for the parametric approaches, mirroring results from the simulations. In the original analysis of the PACE-UP trial, \cite{Harris2017} conduct sensitivity analyses on the impact of using different thresholds on weartime for defining missingness (for example, requiring a minimum of 600 minutes of wear time for a complete day, compared to 540 minutes in the main analyses).  Our analysis additionally reveals the impact of taking an epoch-level perspective on missing data. 

There are a number of important avenues for further work. Firstly, as discussed in \nameref{App}, sensitivity analyses for the MAR assumption are an important consideration in the context of these epoch-level approaches to MI. As participants may be likely to remove their accelerometer during inactive periods, considering the implication of the data being MNAR is important. Secondly, it is worth exploring potential adaptations to the non-parametric approach to improve its performance in the setting where there are entire weeks of data missing. Our simulations studies indicate that replacing entire weeks of data with non-self donors appears to reduce the performance of this approach, potentially because the number of available donors are insufficient in this setting. A hybrid approach which imputes non-wear and sleep-extra non-parametrically and takes advantage of the parametric approach for entire missing weeks would be a potential solution. Thirdly, we have considered the setting where whole-week imputation occurs only at 12 Months, which was the case in the PACE-UP trial. In other studies, there may be participants with missing weeks at baseline included in the study, in which case, donors for whole-week imputation at baseline may be determined by physical activity characteristics obtained at 12 Months. For an extended analysis of the PACE-UP trial including outcomes at 3 Months and 3 Years in addition to baseline and 12 Months, an approach to identifying suitable donors in this more complex setting with multiple time points will need to be identified.\\

While our study has focused on accelerometer studies, there is a growing need to consider missingness at a finer epoch-level for a number of health outcomes in trials, particularly as continuous monitoring participants through digital devices become more common \citep{Dagher2020}. For example, studies that remotely monitor vital signs of people with dementia use a number of sensor and wearable devices that track data in fine intervals of time \citep{David2021}. Our presented framework of classifying missingness and applying MI at the finer epoch-level to establish conclusions about effect of treatment at an aggregate level is likely to be applicable to data from a range of devices. \\ \vspace{10cm}

\section{Appendix}
\subsection{Plots of Vector Magnitude at the Epoch-level} 
\label{Epoch-plots}

\begin{figure}[H]
\center
\includegraphics[scale=0.5]{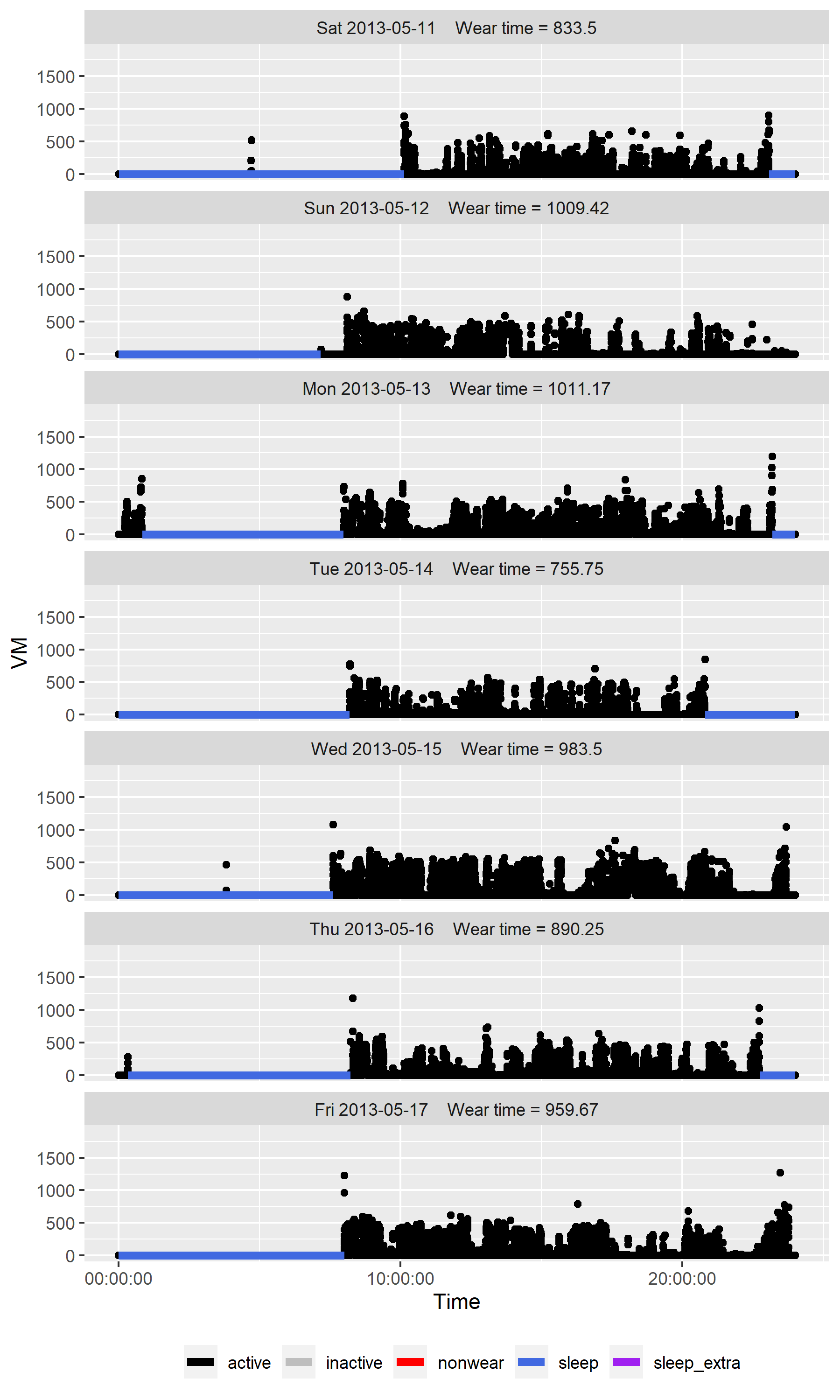}
\caption{An example of 7 days of accelerometer data, where Vector Magnitude at the epoch-level is plotted against time. No missing data is detected. }
\label{}
\end{figure}

\begin{figure}[H]
\center
\includegraphics[scale=0.5]{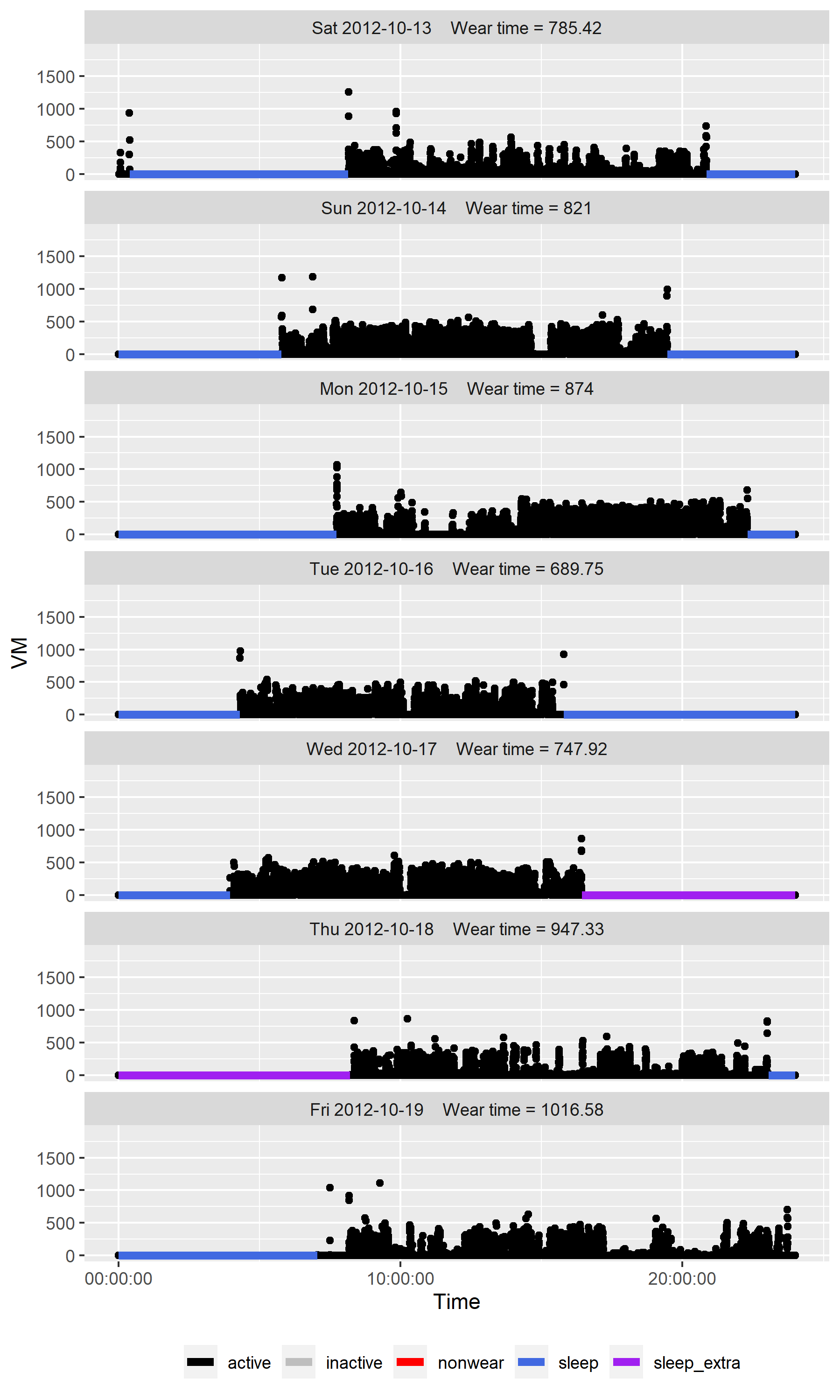}
\caption{An example of 7 days of accelerometer data, where Vector Magnitude at the epoch-level is plotted against time. Sleep-extra is detected between Wednesday and Thursday.}
\label{}
\end{figure}

\begin{figure}[H]
\center
\includegraphics[scale=0.5]{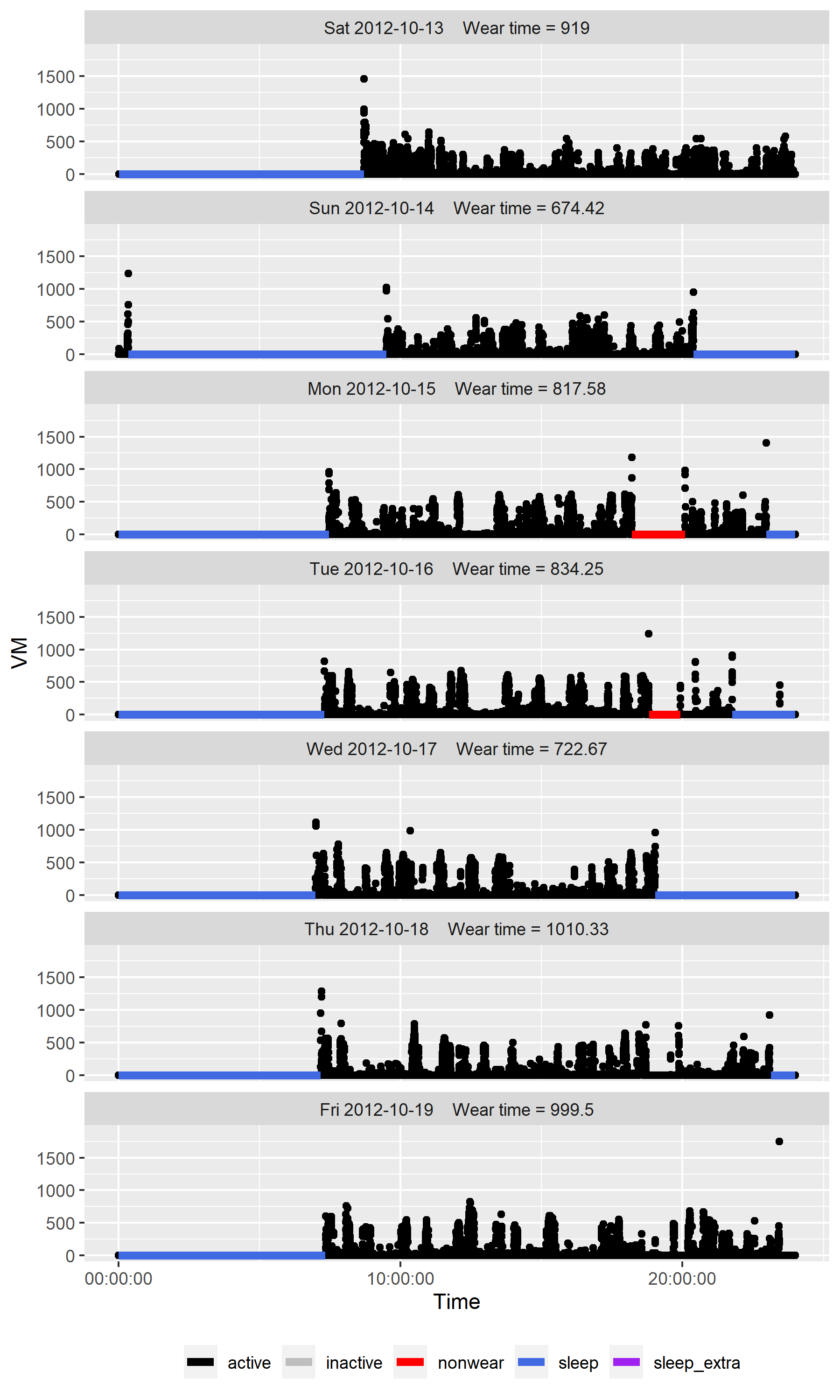}
\caption{An example of 7 days of accelerometer data, where Vector Magnitude at the epoch-level is plotted against time. Non-wear periods are detected on Monday and Tuesday. }
\label{}
\end{figure}

\subsection{PACE-UP trial analysis: Further Results}

\begin{figure}[H]
\center
\includegraphics[scale=0.7]{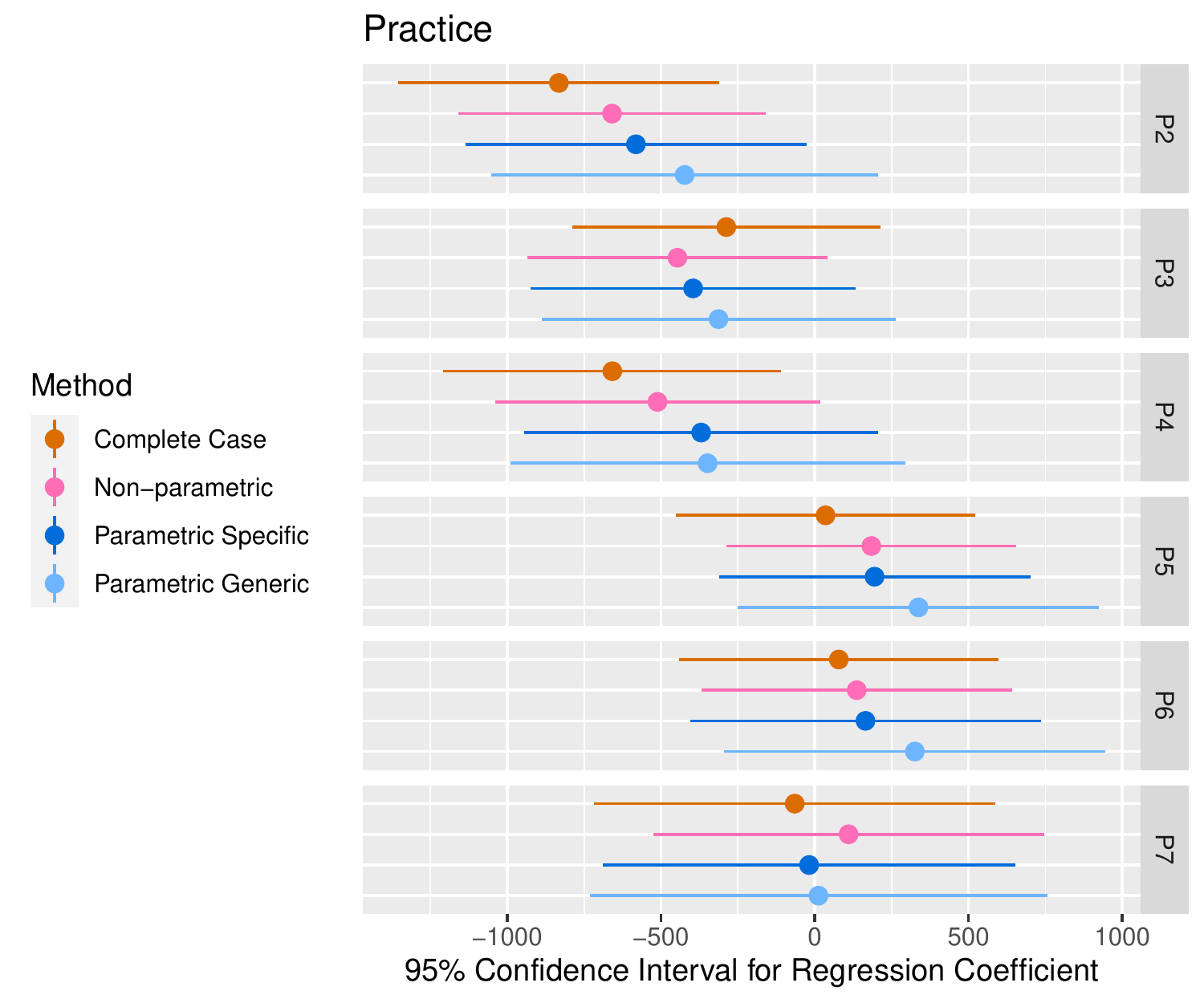}
\caption{Results for the analysis of the PACE-UP trial. For each method of handling missing data, the $95\%$ confidence intervals for the effects of practices are displayed.}
\label{result_practice}
\end{figure}

\begin{figure}[H]
\center
\includegraphics[scale=0.7]{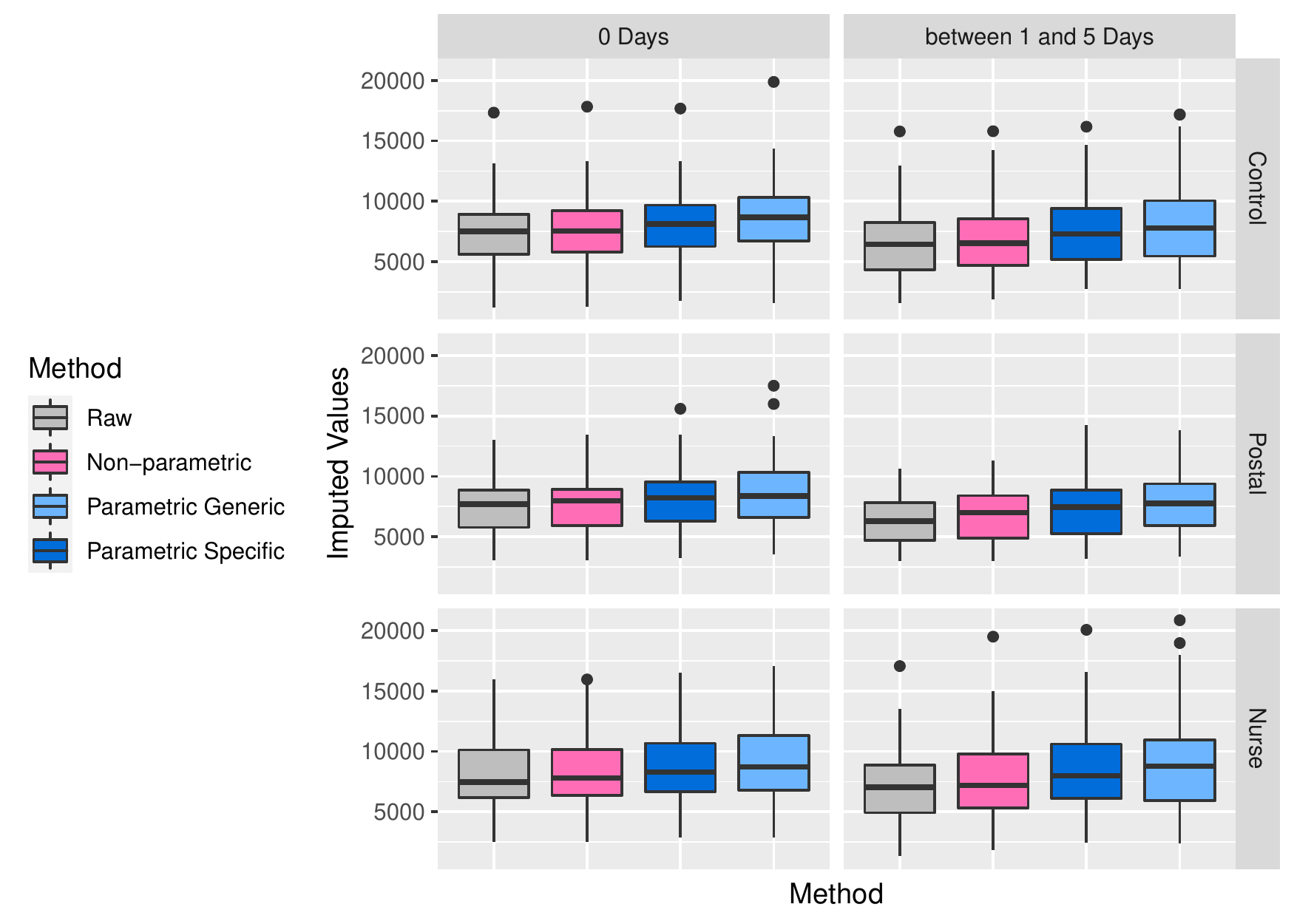}
\caption{Boxplots compare the raw value with the imputed values at baseline under the non-parametric and parametric approaches when there are zero days with weartime $< 540$ minutes (left panel) and between 1 and 5 days with weartime $< 540$ minutes (right panel). Results are shown by treatment arm. }
\label{baseline_boxplots_bygroup}
\end{figure}

\section{Acknowledgements}
The authors would like to thank the South-West London (UK) general practices, their practice nurses who supported the PACE-UP trial, all the patients from these practices who participated, and the Trial Steering Committee. 
Disclaimer: the views and opinions expressed therein are those of the authors and do not necessarily reflect those of the Health Technology Assessment (HTA) programme, National Institute for Health Research (NIHR) National Health Service, or the Department of Health.

\section{Declaration of Conflicting Interests}
The authors have no conflicts of interest.

\section{Funding}
MT is supported by Health Data Research UK, which is funded by the UK Medical Research Council, Engineering and Physical Sciences Research Council, Economic and Social Research Council, Department of Health and Social Care (England), Chief Scientist Office of the Scottish Government Health and Social Care Directorates, Health and Social Care Research and Development Division (Welsh Government), Public Health Agency (Northern Ireland), British Heart Foundation and Wellcome. JC is supported by the Medical Research Council, grant numbers MC UU 12023/21 and MC UU 12023/29. EW is supported by MRC project grants MR/S01442X/1 and MR/R013489/1. TH is supported by the National Institute for Health and Care Research grant numbers NIHR202213, NIHR129074, the National Institute for Health grant number R21 AI156161-01 and Muscular Dystrophy UK grant number 19GROE-PG24-0349-2. The PACE-UP trial was funded by the National Institute for Health Research Health
Technology Assessment (NIHR HTA) Programme
(project number HTA 10/32/02 ISRCTN42122561).

\newpage
\bibliographystyle{SageH}
\bibliography{library.bib}

\begin{thebibliography}{20}
\providecommand{\natexlab}[1]{#1}
\providecommand{\url}[1]{\texttt{#1}}
\providecommand{\urlprefix}{URL }
\expandafter\ifx\csname urlstyle\endcsname\relax
  \providecommand{\doi}[1]{DOI:\discretionary{}{}{}#1}\else
  \providecommand{\doi}{DOI:\discretionary{}{}{}\begingroup
  \urlstyle{rm}\Url}\fi

\bibitem[{Andridge and Little(2010)}]{Andridge2010}
Andridge RR and Little RJ (2010) {A review of hot deck imputation for survey
  non-response}.
\newblock \emph{International Statistical Review} 78(1): 40--64.
\newblock \doi{10.1111/j.1751-5823.2010.00103.x}.

\bibitem[{Butera et~al.(2019)Butera, Li, Evenson, Di, Buchner, LaMonte, LaCroix
  and Herring}]{Butera2019}
Butera NM, Li S, Evenson KR, Di C, Buchner DM, LaMonte MJ, LaCroix AZ and
  Herring A (2019) Hot deck multiple imputation for handling missing
  accelerometer data.
\newblock \emph{Stat. Biosci.} 11(2): 422--448.

\bibitem[{Cameron et~al.(2017)Cameron, Godino, Nichols, Wing, Hill and
  Patrick}]{Cameron2017}
Cameron N, Godino J, Nichols JF, Wing D, Hill L and Patrick K (2017)
  {Associations between physical activity and BMI, body fatness, and visceral
  adiposity in overweight or obese Latino and non-Latino adults}.
\newblock \emph{International Journal of Obesity} 41(6): 873--877.

\bibitem[{Carpenter and Kenward(2012)}]{Carpenter2012}
Carpenter J and Kenward M (2012) \emph{Multiple Imputation and its
  Application}.
\newblock {John Wiley \& Sons, Ltd}.
\newblock ISBN 9781119942283.
\newblock \doi{10.1002/9781119942283}.

\bibitem[{Choi et~al.(2011)Choi, Liu, Matthews and Buchowski}]{Choi2011}
Choi L, Liu Z, Matthews CE and Buchowski MS (2011) Validation of accelerometer
  wear and nonwear time classification algorithm.
\newblock \emph{Med. Sci. Sports Exerc.} 43(2): 357--364.

\bibitem[{Cro et~al.(2020)Cro, Morris, Kenward and Carpenter}]{Cro2020}
Cro S, Morris TP, Kenward MG and Carpenter JR (2020) Sensitivity analysis for
  clinical trials with missing continuous outcome data using controlled
  multiple imputation: A practical guide.
\newblock \emph{Statistics in Medicine} 39(21): 2815--2842.
\newblock
  \urlprefix\url{https://onlinelibrary.wiley.com/doi/abs/10.1002/sim.8569}.

\bibitem[{Dagher et~al.(2020)Dagher, Shi, Zhao and Marrouche}]{Dagher2020}
Dagher L, Shi H, Zhao Y and Marrouche NF (2020) Wearables in cardiology: Here
  to stay.
\newblock \emph{Heart Rhythm} 17(5, Part B): 889--895.
\newblock \doi{https://doi.org/10.1016/j.hrthm.2020.02.023}.
\newblock
  \urlprefix\url{https://www.sciencedirect.com/science/article/pii/S1547527120301740}.
\newblock Digital Health Special Issue.

\bibitem[{David et~al.(2021)David, Barnaghi, Nilforooshan, Rostill, Soreq,
  Sharp and Scott}]{David2021}
David M, Barnaghi P, Nilforooshan R, Rostill H, Soreq E, Sharp DJ and Scott G
  (2021) {Home monitoring of vital signs and generation of alerts in a cohort
  of people living with dementia}.
\newblock \emph{Alzheimer's \& dementia : the journal of the Alzheimer's
  Association} 17: e055151.
\newblock \doi{10.1002/alz.055151}.

\bibitem[{{De Craemer} et~al.(2016){De Craemer}, {De Decker}, Verloigne, {De
  Bourdeaudhuij}, Manios and Cardon}]{DeCraemer2016}
{De Craemer} M, {De Decker} E, Verloigne M, {De Bourdeaudhuij} I, Manios Y and
  Cardon G (2016) {The effect of a cluster randomised control trial on
  objectively measured sedentary time and parental reports of time spent in
  sedentary activities in Belgian preschoolers: The ToyBox-study}.
\newblock \emph{International Journal of Behavioral Nutrition and Physical
  Activity} 13(1): 1--17.
\newblock \urlprefix\url{http://dx.doi.org/10.1186/s12966-015-0325-y}.

\bibitem[{Evenson and Terry(2009)}]{Evenson2009}
Evenson K and Terry J (2009) Assessment of differing definitions of
  accelerometer nonwear time.
\newblock \emph{Research quarterly for exercise and sport} 80: 355--62.

\bibitem[{Goode et~al.(2015)Goode, Winkler, Reeves and Eakin}]{Goode2015}
Goode AD, Winkler EAH, Reeves MM and Eakin EG (2015) {Relationship between
  intervention dose and outcomes in living well with diabetes--a randomized
  trial of a telephone-delivered lifestyle-based weight loss intervention.}
\newblock \emph{American journal of health promotion : AJHP} 30(2): 120--129.
\newblock
  \urlprefix\url{http://ovidsp.ovid.com/ovidweb.cgi?T=JS{\&}PAGE=reference{\&}D=med11{\&}NEWS=N{\&}AN=25372235}.

\bibitem[{Harel and Zhou(2007)}]{Harel2007}
Harel O and Zhou XH (2007) Multiple imputation: review of theory,
  implementation and software.
\newblock \emph{Statistics in Medicine} 26(16): 3057--3077.
\newblock
  \urlprefix\url{https://onlinelibrary.wiley.com/doi/abs/10.1002/sim.2787}.

\bibitem[{Harris et~al.(2018)Harris, Kerry, Limb, Furness, Wahlich, Victor,
  Iliffe, Whincup, Ussher, Ekelund, Fox-Rushby, Ibison, DeWilde, McKay and
  Cook}]{Harris2018}
Harris T, Kerry SM, Limb ES, Furness C, Wahlich C, Victor CR, Iliffe S, Whincup
  PH, Ussher M, Ekelund U, Fox-Rushby J, Ibison J, DeWilde S, McKay C and Cook
  DG (2018) {Physical activity levels in adults and older adults 3-4 years
  after pedometer-based walking interventions: Long-term follow-up of
  participants from two randomised controlled trials in UK primary care.}
\newblock \emph{PLoS medicine} 15(3): e1002526.
\newblock
  \urlprefix\url{http://ovidsp.ovid.com/ovidweb.cgi?T=JS{\&}PAGE=reference{\&}D=medc{\&}NEWS=N{\&}AN=29522529}.

\bibitem[{Harris et~al.(2017)Harris, Kerry, Limb, Victor, Iliffe, Ussher,
  Whincup, Ekelund, Fox-Rushby, Furness, Anokye, Ibison, DeWilde, David,
  Howard, Dale, Smith and Cook}]{Harris2017}
Harris T, Kerry SM, Limb ES, Victor CR, Iliffe S, Ussher M, Whincup PH, Ekelund
  U, Fox-Rushby J, Furness C, Anokye N, Ibison J, DeWilde S, David L, Howard E,
  Dale R, Smith J and Cook DG (2017) Effect of a primary care walking
  intervention with and without nurse support on physical activity levels in
  45- to 75-year-olds: The pedometer and consultation evaluation ({PACE-UP})
  cluster randomised clinical trial.
\newblock \emph{PLOS Medicine} 14(1): 1--19.
\newblock \urlprefix\url{https://doi.org/10.1371/journal.pmed.1002210}.

\bibitem[{Harris et~al.(2015)Harris, Kerry, Victor, Ekelund, Woodcock, Iliffe,
  Whincup, Beighton, Ussher, Limb, David, Brewin, Adams, Rogers and
  Cook}]{Harris2015}
Harris T, Kerry SM, Victor CR, Ekelund U, Woodcock A, Iliffe S, Whincup PH,
  Beighton C, Ussher M, Limb ES, David L, Brewin D, Adams F, Rogers A and Cook
  DG (2015) A primary care nurse-delivered walking intervention in older
  adults: Pace (pedometer accelerometer consultation evaluation)-lift cluster
  randomised controlled trial.
\newblock \emph{PLOS Medicine} 12(2): 1--23.
\newblock \urlprefix\url{https://doi.org/10.1371/journal.pmed.1001783}.

\bibitem[{Ismail et~al.(2019)Ismail, Stahl, Bayley, Twist, Stewart, Ridge,
  Britneff, Ashworth, de~Zoysa, Rundle, Cook, Whincup, Treasure, McCrone,
  Greenough and Winkley}]{Ismail2019}
Ismail K, Stahl D, Bayley A, Twist K, Stewart K, Ridge K, Britneff E, Ashworth
  M, de~Zoysa N, Rundle J, Cook D, Whincup P, Treasure J, McCrone P, Greenough
  A and Winkley K (2019) Enhanced motivational interviewing for reducing weight
  and increasing physical activity in adults with high cardiovascular risk: the
  move it three-arm rct.
\newblock \emph{Health Technol Assess} 23(69).

\bibitem[{{Lee} and Gill(2018)}]{Lee2018}
{Lee} JA and Gill J (2018) {Missing value imputation for physical activity data
  measured by accelerometer}.
\newblock \emph{Statistical Methods in Medical Research} 27(2): 490--506.

\bibitem[{Leeger-Aschmann et~al.(2019)Leeger-Aschmann, Schmutz, Zysset,
  Kakebeeke, Messerli-B\"{u}rgy, St\"{u}lb, Arhab, Meyer, Munsch, Jenni, Puder
  and Kriemler}]{LeegerAschmann2019}
Leeger-Aschmann CS, Schmutz EA, Zysset AE, Kakebeeke TH, Messerli-B\"{u}rgy N,
  St\"{u}lb K, Arhab A, Meyer AH, Munsch S, Jenni OG, Puder JJ and Kriemler S
  (2019) Accelerometer-derived physical activity estimation in preschoolers
  {\textendash} comparison of cut-point sets incorporating the vector magnitude
  vs the vertical axis.
\newblock \emph{{BMC} Public Health} 19(1).
\newblock \doi{10.1186/s12889-019-6837-7}.
\newblock \urlprefix\url{https://doi.org/10.1186/s12889-019-6837-7}.

\bibitem[{Rubin(1976)}]{Rubin1976}
Rubin DB (1976) Inference and missing data.
\newblock \emph{Biometrika} 63(3): 581--592.
\newblock \urlprefix\url{http://www.jstor.org/stable/2335739}.

\bibitem[{Tackney et~al.(2021)Tackney, Cook, Stahl, Ismail, Williamson and
  Carpenter}]{Tackney2021}
Tackney MS, Cook DG, Stahl D, Ismail K, Williamson E and Carpenter J (2021) A
  framework for handling missing accelerometer outcome data in trials.
\newblock \emph{Trials} 22(1).
\newblock \doi{10.1186/s13063-021-05284-8}.
\newblock \urlprefix\url{https://doi.org/10.1186%2Fs13063-021-05284-8}.

\end{thebibliography}

\end{document}